\documentclass[journal]{IEEEtran}
\usepackage{amsmath,amsfonts}
\usepackage{algorithmic}
\usepackage{array}
\usepackage{textcomp}
\usepackage{stfloats}
\usepackage{url}
\usepackage{verbatim}
\usepackage{graphicx}

\usepackage{multirow}
\usepackage{float}
\IEEEoverridecommandlockouts
\usepackage{cite}
\usepackage{xcolor}
\usepackage{xcolor}
\usepackage{epstopdf}
\usepackage{lettrine}
\usepackage{bm}
\usepackage{booktabs}
\usepackage{multirow}
\usepackage{makecell}
\usepackage{color}
\usepackage{subfigure}
\usepackage{multirow}
\usepackage{booktabs}
\usepackage{algorithm}
\usepackage{makecell}
\usepackage{xpatch}
\begin{document}

\title{Spectrum Prediction With Deep 3D Pyramid \\Vision Transformer Learning}

\author{Guangliang~Pan,~\IEEEmembership{Graduate~Student~Member,~IEEE,}
	Qihui~Wu,~\IEEEmembership{Fellow,~IEEE,}
	Bo~Zhou,~\IEEEmembership{Member,~IEEE,}
	Jie~Li,~\IEEEmembership{Member,~IEEE,}
	Wei~Wang,~\IEEEmembership{Member,~IEEE,}
	Guoru~Ding,
	and David~K. Y.~Yau,~\IEEEmembership{Senior~Member,~IEEE}		
\thanks{This work was supported in part by the National Natural Science Foundation of China (NNSFC) under Grant 62231015, in part by the National Research Foundation, Singapore and the Infocomm Media Development Authority, Singapore under the Future Communications Research and Development Programme, award number FCP-SUTD-RG-2021-004 (research was performed while G. Pan was a visiting PhD student in the SUTD Future Communications Lab), and in part by the NNSFC under Grant 62201255. This paper has submitted in part at IEEE WCNC 2025 \cite{sp}. (\textit{Corresponding author: Bo Zhou.})}
\thanks{G. Pan, Q. Wu, B. Zhou, J. Li, and W. Wang are with the College of Electronic and Information Engineering, Nanjing University of Aeronautics and Astronautics, Nanjing, 211106, China (e-mail: \{glpan2020, wuqihui, b.zhou, lijie\_evelyn, wei\_wang\}@nuaa.edu.cn).}
\thanks{G. Ding is with the College of Communications Engineering, Army Engineering University, Nanjing 210007, China (e-mail: guoru\_ding@yeah.net).}
\thanks{D. Yau is with the Pillar of Information Systems Technology and Design, Singapore University of Technology and Design, Singapore 487372 (e-mail: david\_yau@sutd.edu.sg).}
}



\maketitle

\begin{abstract}
In this paper, we propose a deep learning (DL)-based task-driven spectrum prediction framework, named DeepSPred. The DeepSPred comprises a feature encoder and a task predictor, where the encoder extracts spectrum usage pattern features, and the predictor configures different networks according to the task requirements to predict future spectrum. Based on the DeepSPred, we first propose a novel 3D spectrum prediction method combining a flow processing strategy with 3D vision Transformer (ViT, i.e., Swin) and a pyramid to serve possible applications such as spectrum monitoring task, named 3D-SwinSTB. 3D-SwinSTB unique \textit{3D Patch Merging ViT-to-3D ViT Patch Expanding} and pyramid designs help the model accurately learn the potential correlation of the evolution of the spectrogram over time. Then, we propose a novel spectrum occupancy rate (SOR) method by redesigning a predictor consisting exclusively of 3D convolutional and linear layers to serve possible applications such as dynamic spectrum access (DSA) task, named 3D-SwinLinear. Unlike the 3D-SwinSTB output spectrogram, 3D-SwinLinear projects the spectrogram directly as the SOR. Finally, we employ transfer learning (TL) to ensure the applicability of our two methods to diverse spectrum services. The results show that our 3D-SwinSTB outperforms recent benchmarks by more than 5\%, while our 3D-SwinLinear achieves a 90\% accuracy, with a performance improvement exceeding 10\%.    
\end{abstract}


\section{Introduction}
\lettrine[lines=2]{T}{he} scarcity of the radio-frequency (RF) spectrum is a critical concern within the domain of wireless communications, primarily due to the constrained availability of frequency bands for utilization \cite{10106497}. As the demand for wireless services grow rapidly, the limited availability of the RF spectrum has emerged as a substantial challenge for the wireless industry, leading to congestion and diminished service quality. This predicament is further exacerbated by the exponential surge in data traffic propelled by the Internet of Things (IoT), particularly smartphones and cloud-based services \cite{9424395}.

Cognitive radio (CR) technology has emerged as a promising solution to mitigate the impact of spectrum scarcity because it has the potential to revolutionize the way we use and manage the RF spectrum. Traditional spectrum allocation methods, which are based on static and exclusive allocation of frequencies to specific spectrum users, have led to inefficient use of the spectrum, limiting its availability and accessibility for new and emerging applications. CR addresses this issue by allowing for dynamic spectrum access (DSA), making better use of the available spectrum and improving overall spectrum utilization efficiency \cite{huang2019dynamic}. As a compensation technique, spectrum prediction can efficiently improve the performance of a CR network (CRN) for tasks such as spectrum monitoring and spectrum access \cite{7955907, 9241879}. Specifically, spectrum prediction in spectrum monitoring can provide possible future anomalies to help spectrum managers optimize spectrum allocation and sharing protocols. Furthermore, spectrum prediction can provide future spectrum occupancy to help secondary users (SUs) identify free frequency bands and access them. 

However, spectrum entities in a CRN usually require different types of spectrum prediction information depending on the task they perform. For example, in a spectrum monitoring task, \textit{spectrogram} can provide a visual representation of anomalies, which allows a spectrum manager to quickly optimize spectrum usage. In a spectrum access task, \textit{spectrum occupancy rate} (SOR) \cite{6933906} can provide intuitive time-frequency occupancy information, which is sent for SUs to help them with fast DSA. Therefore, when facing multiple spectrum tasks, it is a crucial problem to configure the corresponding type of spectrum information according to different tasks. To the best of our knowledge, existing spectrum prediction work (see Section \ref{sec2}) does not take this issue into account. Providing accurate spectrum prediction information for each spectrum task depends on the advanced nature of the prediction method adopted. Recently, deep learning (DL)-based approaches have gained popularity in spectrum prediction due to their capability to handle nonlinear data effectively \cite{8456449, shawel2019convolutional, yu2020spectrum, 9296309, 9664805}. Data-driven DL-methods generally do not rely on prior information and can automatically extract relevant features from the spectrum data. However, despite the progress made by DL-based spectrum prediction methods, several challenges remain. Existing DL-based approaches commonly employ recurrence and convolution networks to construct hybrid networks that can capture the temporal, spectral, and spatial correlations of the spectrum \cite{yu2020spectrum, 9296309, 9664805}. However, recurrence networks are limited in their ability to capture long-term dependencies and can only learn usage pattern between adjacent time steps. The convolution networks have limitations in capturing global usage pattern of frequency band as they primarily focus on local usage pattern defined by the convolution kernel. 

In this paper, we propose a spectrum prediction framework and two spectrum prediction methods to address the aforementioned challenges. Specifically, we first propose a DL-based task-driven spectrum prediction framework, named DeepSPred. The DeepSPred consists of a \textit{feature encoder} and a \textit{task predictor}, where the feature encoder extracts hidden usage patterns in historical spectrum data, and the task predictor configures different networks according to the task requirements (possible applications such as spectrum monitoring and DSA tasks are considered in this work) and infers prediction information corresponding to task type based on extracted pattern features. Based on the proposed DeepSPred, we introduce a novel 3D spectrum prediction method to serve the spectrum monitoring task, namely 3D-SwinSTB, by adapting a flow processing strategy with 3D Swin Transformer \cite{liu2021swin} and a pyramid structure \cite{ronneberger2015u}. The 3D-SwinSTB's flow processing strategy uses a 3D vision self-attention mechanism to assign different attention weights to all temporal positions in the spectrogram series, capturing long-term spatiotemporal dependencies across successive time steps, rather than limiting the focus to short-term dependencies among adjacent time steps like recurrence networks. Compared to convolution networks, 3D-SwinSTB can capture multi-scale features and fuse local and global spectrum usage patterns thanks to flow processing strategy's hierarchical architecture and the alternating computation of self-attention within \textit{3D windows} (capture local patterns) and \textit{3D shifted windows} (capture global patterns). Furthermore, 3D-SwinSTB's pyramid assists flow processing strategy to combat the loss and increased computational complexity incurred by the propagation of features layer by layer. Then, based on the proposed DeepSPred, we develop a novel SOR prediction method to serve the DSA task equipped with a task predictor composed exclusively of 3D convolutional and linear layers, alongside a 3D-SwinSTB's encoder, named 3D-SwinLinear. This method can directly predict the future SOR values based on the spectrogram series to help SUs make quick decisions for DSA. We also apply transfer learning (TL) to our methods to ensure their adaptability to various spectrum services. Extensive experiments are conducted on three real-world datasets. The results show that 3D-SwinSTB exhibits a notable reduction in average error, amounting to a 5\% improvement compared to existing prediction methods. Notably, this improvement is particularly pronounced when predicting sequences exceeding 6 frames. Meanwhile, the 3D-SwinLinear achieves a 90\% prediction accuracy, surpassing prevailing methods by a performance improvement exceeding 10\%. The application of TL to both of these novel methods proves to be effective.

The remainder of this paper is organized as follows. In Section \ref{sec2}, we discuss the related works. Section \ref{sec3} gives the problem description and the DeepSPred framework. Section \ref{sec4} and Section \ref{sec5} present the design details of 3D-SwinSTB and 3D-SwinLinear, respectively. Subsequently, we discuss extensive simulation results in Section \ref{sec6}. Finally, the concluding remarks are summarized in Section \ref{sec7}.

\textit{Notation:} $d$ is a scalar, $\bm{{\rm x}}$ is a vector, and $\bm{{\rm X}}$ is a RGB matrix. $\bm{{\rm X}}_{1:T}$ is a 3-order tensor consisting of $T$ square RGB matrices. $\mathbb{C}$ stands for a complex number field. $\mathbb{R}$ stands for a real number field. $|\cdot|^2$ stands for the squared modulus operation. $p(\cdot|\cdot)$ is a conditional probability. $(\cdot)^T$ is defined as the transpose of a matrix. $\lceil \cdot \rceil$ stands for the rounded up of a scalar. $\mathcal{L}(\cdot)$ stands for a loss function. $\|\cdot\|_2$ stands for the $l_2$-norm, which calculates the square root of the sum of the square of all matrix's elements.
     
\section{Related Work}\label{sec2}
This section reviews the related work on the autoregressive (AR), traditional machine learning (ML), recurrence and convolution networks, and Transformer for spectrum prediction.
\subsection{AR-Based Methods}
In the AR modeling, the classical AR and moving average (MA) techniques were utilized for spectrum prediction \cite{9163307}. Subsequently, the autoregressive moving average (ARMA) \cite{liu2018filter} and autoregressive integrated moving average (ARIMA) \cite{8930636} were proposed to further enhance the prediction accuracy. However, these models can be subject to limitations due to assumptions and simplifications made about the underlying system \cite{7902233}. For instance, certain models may assume a Gaussian distribution of spectrum usage, whereas in reality, it may exhibit greater complexity. Additionally, these models are typically effective for short-term predictions within a limited time range but struggle to capture the long-term dependencies and trends of nonlinear spectrum data.
\subsection{Traditional ML-Based Methods}
Traditional ML methods, including support vector machine (SVM), support vector regression (SVR), and hidden Markov model (HMM), were employed for spectrum prediction \cite{8031332, yu2008frequency, eltom2018cooperative, luo2021temporal}. Hidden bivariate Markov model (HBMM) and higher-order HMM were further developed to enhance the capabilities of HMMs \cite{8031332}. Bayesian approach \cite{xing2013channel} was also employed for spectrum prediction. However, SVMs require careful selection and tuning of model parameters, such as the kernel function and regularization parameter, which can be a complex and time-consuming process for a CR system. HMMs have limitations in terms of limited context modeling, as they rely solely on the previous observation, making it difficult to capture long-term usage pattern dependencies. Bayesian methods have limitations due to their reliance on specific probabilistic distributions, limited access to prior knowledge, and sensitivity to prior distribution selection. 
\subsection{Recurrence and Convolution Networks-Based Methods}
Recurrence and convolution networks are commonly designed as either single networks or hybrid networks for spectrum prediction. In the case of single networks, methodologies such as long short-term memory (LSTM) \cite{lin2020cross} and gated recurrent unit (GRU) \cite{zhao2017machine} are utilized to capture temporal dependencies of the spectrum. In \cite{9625078}, the authors utilized LSTM, \textit{Seq-to-Seq} modeling, and attention for multi-channel multi-step spectrum prediction. Authors in \cite{9339826} utilized deep convolution generative adversarial network (DCGAN) with TL for cross-band prediction. Hybrid learning networks, including deep temporal-spectral residual network (DTS-ResNet) \cite{8456449} and NN-ResNet \cite{9664805}, have been used for spectrum prediction in high-frequency communication and spatiotemporal spectrum load prediction, respectively. These hybrid networks amalgamate convolutional neural network (CNN) and ResNet networks. Further, a learning system proposed in \cite{yu2020spectrum} integrates CNN and GRU for spectrum prediction. Another approach elucidated in \cite{9296309} entails the utilization of predictive recurrent neural network (PredRNN) to acquire knowledge pertaining to various periodic spectrum features. In our prior work \cite{10064355}, we employed stacked autoencoders (SAEs) to extract features in a layer-by-layer manner while reducing dimensionality. Subsequently, we devised a hybrid network comprising fusion convolution and recurrence networks to learn time-frequency-space features. It is noteworthy, however, that LSTM, despite its effectiveness in capturing dependencies among adjacent temporal instances, is incapable of comprehensively encompassing the multidimensional aspects of time-frequency-space features and long-term contextual dependencies. Conversely, hybrid networks such as NN-ResNet can effectively capture spatiotemporal features; nonetheless, their reliance on fixed convolution kernel sizes restricts their ability to capture multi-scale features and global patterns.
\subsection{Transformer-Based Methods}
The Transformer, originally proposed in the field of natural language processing (NLP) \cite{attention}, has found extensive applications in diverse domains for predicting a range of phenomena, including weather patterns, traffic flow, and more \cite{9716741}. In our recent work \cite{10039050}, we have also proposed a long-term spectrum prediction method by integrating the Transformer with the auto-correlation mechanism and series channel-space attention. Nevertheless, this method solely captured temporal dependencies. Recently, the Transformer has been extended to encompass the realm of the computer vision (CV), giving rise to the vision Transformer (ViT) models \cite{9747984, 10105499}. Particularly, a 3D Swin Transformer \cite{liu2021swin} has been effectively applied to video-based recognition tasks, capitalizing on self-attention to capture prolonged spatiotemporal dependencies rather than only capturing temporal dependencies like \cite{10039050}. The Swin's hierarchical design can capture multi-scale features and shifted window design can learn local and global features. Motivated by these advantages, this study investigates the application of the 3D Swin Transformer for spectrum prediction.
\begin{figure}
	\centerline{\includegraphics[width=86mm,height=73mm]{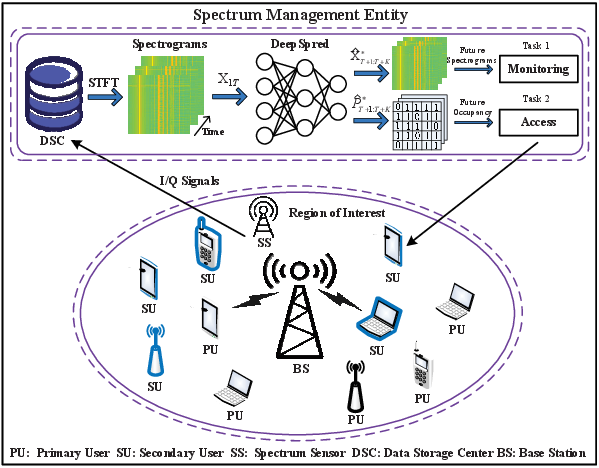}}
	\caption{System model.}
	\label{Fig:system}
\end{figure}
\section{Problem Description and DeepSPred Framework}\label{sec3}
\subsection{Problem Description}\label{3.2}
As shown in Fig. \ref{Fig:system}, we consider a CRN formed by several spectrum users, which consists of primary users (PUs) and SUs, located in a geographical region of interest. The SUs are allowed to dynamically access the free frequency band when the PUs are not using the authorized frequency band. When the PUs resume use of the authorized frequency band, SUs promptly vacate it to avoid interfering with the PUs. A spectrum sensor (SS) is deployed here, which continuously collects the aggregated radio signal $x(t_s)$ from a $F$-bandwidth in the surrounding environment through the antenna. Let $x(t_s)=x_I(t_s)+jx_Q(t_s)$, where $x_I(t_s)$ and $x_Q(t_s)$ are the in-phase (I) and quadrature (Q) signals at time $t_s$. The $x(t_s)$ can originate from aggregated signals sent by all possible wireless devices (such as radio broadcasts, Wi-FI, and cell phone signals) that occupy the spectrum of interest at time $t_s$. The collected signal is transmitted to a data storage center (DSC) in a spectrum management entity deployed on BS. The signal $x(t_s)$ is first transformed by a short-time Fourier transform (STFT):
\begin{equation}\label{Eq_2}
\text{STFT}_x(t_s,f_s)=\int^{+ \infty}_{- \infty} x(\tau)h(\tau-t_s)e^{-j 2 \pi f_s \tau}d\tau,
\end{equation}
where $h(\cdot)$ stands for the window function. Note that $h(\tau-t_s)e^{j 2 \pi f_s \tau}$ has its energy concentrated at time $t_s$ and frequency $f_s$. The STFT can provide both time and frequency information of signal $x(t_s)$, not just temporal information. The signal $x(t_s)$ is then converted into a spectrogram by
\begin{equation}\label{Eq_s}
\begin{aligned}
\text{SPEC}_x(t_s,f_s)=& |\text{STFT}_x(t_s,f_s)|^2 \\
=& |\int^{+ \infty}_{- \infty} x(\tau)h(\tau-t_s)e^{-j 2 \pi f_s \tau}d\tau|^2.
\end{aligned}
\end{equation} 
In this work, we consider two possible task applications of spectrum prediction in spectrum management entity: one task is to monitor the spectrum according to the predicted spectrograms, such as discovering possible anomalies in the future to help the spectrum manager optimize spectrum usage; another task is to send the predicted SOR (for details, see Section \ref{5.1}) to the SUs for making advance decisions regarding dynamic access to idle bands. There are two key problems for these tasks: spectrogram prediction (defined as 3D spectrum prediction) and SOR prediction. Specifically, the spectrum management entity continuously collects length-$T$ historical spectrograms $\bm{{\rm X}}_{1:T} \in \mathbb{R}^{T \times H \times W \times 3}$, $\bm{{\rm X}}_{1:T}=[\bm{{\rm X}}_{1}, \dots, \bm{{\rm X}}_{T}]$, where $\bm{{\rm X}}_{T}$ is $T$th spectrogram, the size of which is $H (\text{height}) \times W (\text{width}) \times 3$ (RGB, the reasons for using it instead of grayscale images are given in Appendix A), $H$ and $W$ correspond to time resolution and frequency resolution, respectively. Based on these spectrograms, the 3D spectrum prediction problem can be represented as
\begin{equation}\label{Eq_3}
\hat{\bm{{\rm X}}}^{*}_{T+1:T+K} = {\rm arg} \mathop {\rm max} \limits_{\bm{{\rm X}}^{*}_{T+1:T+K}} p(\bm{{\rm X}}^{*}_{T+1:T+K}|\bm{{\rm X}}_{1:T}),
\end{equation}
where $\bm{{\rm X}}^{*}_{T+1:T+K} = [\bm{{\rm X}}^{*}_{T+1}, \dots, \bm{{\rm X}}^{*}_{T+K}] \in \mathbb{R}^{K \times H \times W}$ represent the spectrograms in the next $K$ timeslots, and $\hat{\bm{{\rm X}}}^{*}_{T+1:T+K}=[\hat{\bm{{\rm X}}}^{*}_{T+1}, \dots, \hat{\bm{{\rm X}}}^{*}_{T+K}]$. Furthermore, based on these spectrograms, the SOR prediction problem can be represented as
\begin{equation}\label{Eq_33}
\hat{{\rm P}}^{*}_{T+1:T+K} = {\rm arg} \mathop {\rm max} \limits_{{\rm P}^{*}_{T+1:T+K}} p({\rm P}^{*}_{T+1:T+K}|\bm{{\rm X}}_{1:T}),
\end{equation} 
where ${\rm P}^{*}_{T+1:T+K} = [{\rm P}^{*}_{T+1}, \dots, {\rm P}^{*}_{T+K}] \in \mathbb{R}^{K \times 1}$ represent the SOR in the next $K$ timeslots, and $\hat{{\rm P}}^{*}_{T+1:T+K}=[\hat{{\rm P}}^{*}_{T+1}, \dots, \hat{{\rm P}}^{*}_{T+K}]$. From (\ref{Eq_3}) and (\ref{Eq_33}), two different types of spectrum data based on $\bm{{\rm X}}_{1:T}$ need to be predicted. 
\begin{figure}
	\centerline{\includegraphics[width=88mm,height=48mm]{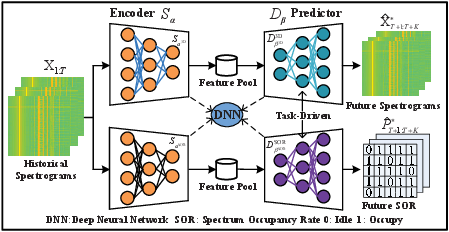}}
	\caption{The DeepSPred framework.}
	\label{Fig:framework}
\end{figure}
 
\subsection{DeepSPred Framework}\label{3.1}
To address the problems (\ref{Eq_3}) and (\ref{Eq_33}), we propose a DL-based task-driven spectrum prediction framework, named DeepSPred, as shown in Fig. \ref{Fig:framework}. From Fig. \ref{Fig:framework}, this framework adopts an \textit{encoder-decoder} structure, which includes two parts: a feature encoder, defined as $S_{\bm{\alpha}}(\cdot)$ and a task predictor, defined as $D_{\bm{\beta}}(\cdot)$. The feature encoder $S_{\bm{\alpha}}(\cdot)$ and task predictor $D_{\bm{\beta}}(\cdot)$ are composed of deep neural networks (DNN) with the encoder parameter set $\bm{\alpha}$ and DNN with the predictor parameter set $\bm{\beta}$, respectively. 
	
The input of the DeepSPred is length-$T$ historical spectrograms $\bm{{\rm X}}_{1:T}$. The output of the DeepSPred includes future spectrograms $\hat{\bm{{\rm X}}}^{*}_{T+1:T+K}$ and SOR $\hat{{\rm P}}^{*}_{T+1:T+K}$. The $S_{\bm{\alpha}}(\cdot)$ extracts hidden usage patterns in historical spectrograms $\bm{{\rm X}}_{1:T}$, and the $D_{\bm{\beta}}(\cdot)$ configures different networks according to the task requirements (i.e., spectrum monitoring task and DSA task) and infers future $\hat{\bm{{\rm X}}}^{*}_{T+1:T+K}$ and $\hat{{\rm P}}^{*}_{T+1:T+K}$ based on extracted pattern features. The processes of encoding $\bm{{\rm X}}_{1:T}$ and predicting $\hat{\bm{{\rm X}}}^{*}_{T+1:T+K}$ can be represented as  
\begin{equation}\label{Eq_sp}
\hat{\bm{{\rm X}}}^{*}_{T+1:T+K} = D^{\text{3D}}_{\bm{\beta}^{\text{3D}}}(S_{\bm{\alpha}^{\text{3D}}}(\bm{{\rm X}}_{1:T})),
\end{equation}
where $D^{\text{3D}}_{\bm{\beta}^{\text{3D}}}(\cdot)$ stands for a configured task predictor with the parameter set $\bm{\beta}^{\text{3D}}$ to solve problem (\ref{Eq_3}). Furthermore, the processes of encoding $\bm{{\rm X}}_{1:T}$ and predicting $\hat{{\rm P}}^{*}_{T+1:T+K}$ can be represented as  
\begin{equation}\label{Eq_ssp}
\hat{{\rm P}}^{*}_{T+1:T+K} = D^{\text{SOR}}_{\bm{\beta}^{\text{SOR}}}(S_{\bm{\alpha}^{\text{SOR}}}(\bm{{\rm X}}_{1:T})),
\end{equation}
where $D^{\text{SOR}}_{\bm{\beta}^{\text{SOR}}}(\cdot)$ stands for a configured task predictor with the parameter set ${\bm{\beta}}^{\text{SOR}}$ to solve problem (\ref{Eq_33}). The different parameters ($\bm{\alpha}^{\text{3D}}$, $\bm{\alpha}^{\text{SOR}}$) are adapted in (\ref{Eq_sp}) and (\ref{Eq_ssp}) with the same encoder $S_{\bm{\alpha}}(\cdot)$ structure.
\begin{algorithm}[t]
	\caption{DeepSPred Framework Training Algorithm} 
	\label{alg:Framwork} 
	\hspace*{0.04in}{\textbf{Initialization}: Initial parameters $\{\bm{\alpha}^{\text{3D}}, \bm{\beta}^{\text{3D}},\bm{\alpha}^{\text{SOR}}, \bm{\beta}^{\text{SOR}}\}$.}
	
	\begin{algorithmic}[1]
		\STATE \textbf{Input}: The historical spectrograms $\bm{{\rm X}}_{1:T}$;  
		\STATE \textbf{while} Stop criterion is not met \textbf{do}
		\STATE \quad $ \hat{\bm{{\rm X}}}^{*}_{T+1:T+K} = D^{\text{3D}}_{\bm{\beta}^{\text{3D}}}(S_{\bm{\alpha}^{\text{3D}}}(\bm{{\rm X}}_{1:T}))$;\\
		\STATE \quad Compute loss function;\\
		\STATE \quad Train $\{\bm{\alpha}^{\text{3D}}, \bm{\beta}^{\text{3D}}\}$ by optimizer;\\
		\STATE \textbf{end while}
		\STATE \textbf{while} Stop criterion is not met \textbf{do}
		\STATE \quad $\hat{{\rm P}}^{*}_{T+1:T+K} = D^{\text{SOR}}_{\bm{\beta}^{\text{SOR}}}(S_{\bm{\alpha}^{\text{SOR}}}(\bm{{\rm X}}_{1:T}))$;\\
		\STATE \quad Compute loss function;\\
		\STATE \quad Train $\{\bm{\alpha}^{\text{SOR}}, \bm{\beta}^{\text{SOR}}\}$ by optimizer;\\
		\STATE \textbf{end while}
		\STATE Obtain updated parameters $\{{{\bm{\alpha}}^{*}}^{\text{3D}}, {{\bm{\beta}}^{*}}^{\text{3D}}, {{\bm{\alpha}}^{*}}^{\text{SOR}}, {{\bm{\beta}}^{*}}^{\text{SOR}}\}$;\\
		\STATE \textbf{Output}: The trained framework\\ 
		\STATE $\{{D}^{\text{3D}}_{{\bm{\beta}^{*}}^{\text{3D}}}(S_{{\bm{\alpha}^{*}}^{\text{3D}}}(\cdot)), {D}^{\text{SOR}}_{{\bm{\beta}^{*}}^{\text{SOR}}}(S_{{\bm{\alpha}^{*}}^{\text{SOR}}}(\cdot))\}$.	
	\end{algorithmic}
\end{algorithm}

The DeepSPred needs to be trained before the prediction. The specific training process is shown in Algorithm \ref{alg:Framwork}. In Algorithm \ref{alg:Framwork}, the task predictor $D_{\bm{\beta}}(\cdot)$ is set to $D^{\text{3D}}_{\bm{\beta}^{\text{3D}}}(\cdot)$ if a spectrum monitoring task is performed. The $D_{\bm{\beta}}(\cdot)$ is set to $D^{\text{SOR}}_{\bm{\beta}^{\text{SOR}}}(\cdot)$ if a spectrum access task is performed. Note that these two tasks are trained separately during the training phase. The stopping criterion is that the loss function (for details, see (\ref{loss})) between the predicted result (i.e., $\hat{\bm{{\rm X}}}^{*}_{T+1:T+K}$ or $\hat{{\rm P}}^{*}_{T+1:T+K}$) and the true result (i.e., $\bm{{\rm X}}^{*}_{T+1:T+K}$ or ${\rm P}^{*}_{T+1:T+K}$) reaches convergence, where the convergence condition is that the loss decreases by less than $\vartheta_{\text{per}}$ \% in $n_{\text{ep}}$ consecutive epochs. Once trained, the trained framework $D_{{\bm{\beta}}^{*}}(S_{{\bm{\alpha}}^{*}}(\cdot))$ is used to predict $\hat{\bm{{\rm X}}}^{*}_{T+1:T+K}$ and $\hat{{\rm P}}^{*}_{T+1:T+K}$. Specifically,
\begin{equation}
\hat{\bm{{\rm X}}}^{*}_{T+1:T+K} = {D}^{\text{3D}}_{{\bm{\beta}^{*}}^{\text{3D}}}(S_{{\bm{\alpha}^{*}}^{\text{3D}}}(\bm{{\rm X}}_{1:T})),
\end{equation}
\begin{equation}
\hat{{\rm P}}^{*}_{T+1:T+K} = {D}^{\text{SOR}}_{{\bm{\beta}^{*}}^{\text{SOR}}}(S_{{\bm{\alpha}^{*}}^{\text{SOR}}}(\bm{{\rm X}}_{1:T})),
\end{equation}
where $S_{{\bm{\alpha}^{*}}^{\text{3D}}}(\cdot)$ and ${D}^{\text{3D}}_{{\bm{\beta}^{*}}^{\text{3D}}}(\cdot)$ stand for the trained 3D spectrum encoder with the updated parameter set ${\bm{\alpha}^{*}}^{\text{3D}}$ and predictor with the updated parameter set ${\bm{\beta}^{*}}^{\text{3D}}$, respectively, and $S_{{\bm{\alpha}^{*}}^{\text{SOR}}}(\cdot)$ and ${D}^{\text{SOR}}_{{\bm{\beta}^{*}}^{\text{SOR}}}(\cdot)$ stand for the trained SOR encoder with the updated parameter set ${\bm{\alpha}^{*}}^{\text{SOR}}$ and predictor with the updated parameter set ${\bm{\beta}^{*}}^{\text{SOR}}$, respectively.
\begin{figure*}
	\centerline{\includegraphics[width=182mm,height=50.5mm]{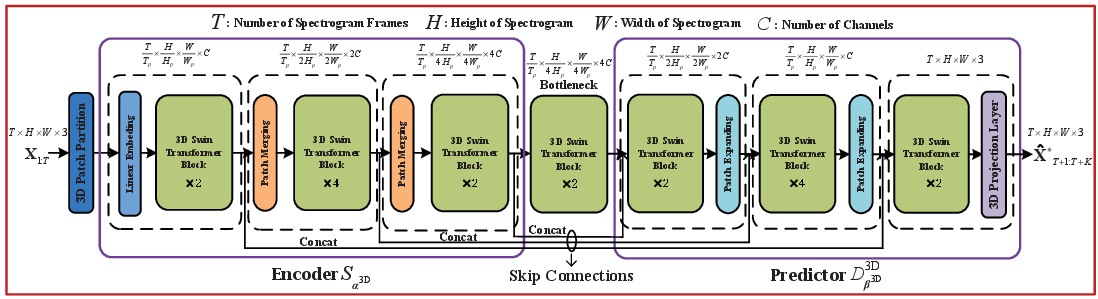}}
	\caption{The structure of the 3D-SwinSTB. Here, $T_p$, $H_p$, and $W_p$ represent the frame number, height, and width of each 3D patch, respectively.}
	\label{Fig:3D-SwinM}
\end{figure*}

Next, we discuss the advantages of the DeepSPred from three aspects: problem modeling, framework structure, and learning performance. Specifically, 
\begin{itemize}
	\item DeepSPred employs a seq-to-seq modeling, which can effectively capture long-range spatiotemporal dependencies. It accommodates input and prediction sequences of varying lengths, as demonstrated in our prior work \cite{10039050}. Further, DeepSPred can provide various types of prediction information to meet downstream task requirements.
	\item The encoder and predictor can be independently designed according to the task needs. This flexibility allows DeepSPred to be applied to a wide range of spectrum prediction tasks. Further, DeepSPred leverages this flexibility to configure advanced components such as the 3D-ViT used in this paper, collaboratively designed with the encoder-decoder to achieve high-precision predictions.
	\item The encoder extracts abstract features from spectrograms, which the predictor uses to generate future spectrograms and SOR. This abstraction helps the model generalize better to unseen spectrum data and capture the underlying structure. Further, DeepSPred uses a parallel prediction method to quickly provide users with spectrum information, thereby reducing decision latency. 
\end{itemize}

It is evident from (\ref{Eq_sp}) and (\ref{Eq_ssp}) that for precise prediction results, four components $S_{\bm{\alpha}^{\text{3D}}}(\cdot)$, $S_{\bm{\alpha}^{\text{SOR}}}(\cdot)$, $D^{\text{3D}}_{\bm{\beta}^{\text{3D}}}(\cdot)$, and $D^{\text{SOR}}_{\bm{\beta}^{\text{SOR}}}(\cdot)$ need to be individually designed. The design specifics of these four components will be elaborated in the subsequent sections using two spectrum prediction methods.

\section{Proposed 3D-SwinSTB Method Design \\ for 3D Spectrum Prediction}\label{sec4}
In this section, we propose a 3D spectrum prediction method named as 3D-SwinSTB to solve (\ref{Eq_sp}). The overall architecture of the proposed 3D-SwinSTB is shown in Fig. \ref{Fig:3D-SwinM}. Specifically, the 3D-SwinSTB is designed based on the DeepSPred, which consists of an encoder $S_{\bm{\alpha}^{\text{3D}}}$, a predictor $D^{\text{3D}}_{\bm{\beta}^{\text{3D}}}$, and a pyramid structure. The $S_{\bm{\alpha}^{\text{3D}}}$ consists of \textit{3D Patch Partition}, \textit{Linear Embedding}, \textit{3D Swin Transformer Block}, and \textit{Patch Merging}. The $D^{\text{3D}}_{\bm{\beta}^{\text{3D}}}$ consists of \textit{Patch Expanding}, \textit{3D Swin Transformer Block}, and \textit{3D Projectiion Layer}. The pyramid structure consists of a bottleneck layer and a skip connection. The proposed 3D-SwinSTB is different from that of traditional ViT \cite{liu2021swin} in two aspects: (i) we propose a novel \textit{3D Patch Merging ViT-to-3D ViT Patch Expanding} symmetric flow processing strategy to learn the spectrum usage pattern and spatiotemporal dependence at different frequency bands in the spectrogram series, while traditional ViT only stacks multiple ViT blocks together; (ii) we design a pyramid structure integrated with flow processing strategy to combat the loss and increased computational complexity incurred by the propagation of usage pattern features layer by layer, while traditional ViT only relies on increasing the number of blocks to counteract the loss. Below, we introduce them in detail.  

\subsection{Encoder, Predictor, and Pyramid}\label{4.1}
\textit{Encoder $S_{\bm{\alpha}^{\text{3D}}}$:} As shown in Fig. \ref{Fig:3D-SwinM}, the historical spectrograms $\bm{{\rm X}}_{1:T} \in \mathbb{R}^{T \times H \times W \times 3}$ ($T$-frame $H \times W \times 3$ RGB pixels) are first split into non-overlapping 3D patches by the 3D patch partition. We treat each 3D patch as a token consisting multidimensional spectrum features. Then, the features of each token are fed into a linear embedding layer, which is projected onto an arbitrary dimension (the number of dimensions is denoted as $C$). The transformed tokens are entered into several 3D Swin Transformer blocks and patch merging layers in turn to generate the high-level usage pattern representations. The process of encoding can be given by
\begin{equation}\label{eq5}
\left\{
\begin{aligned}
\mathcal{X}_{\text{en}} =&\ {\small \text{\sffamily{3DPatchPar}}}(\bm{{\rm X}}_{1:T}), \\
\mathcal{S}^{1}_{\text{en}} =&\ {\small \text{\sffamily{3DSwinTrans}}}(\small{\text{\sffamily{LinearEm}}}(\mathcal{X}_{\text{en}})), \\
\mathcal{S}^{2}_{\text{en}} =&\ {\small \text{\sffamily{3DSwinTrans}}}({\small{\text{\sffamily{PatchMer}}}}(\mathcal{S}^{1}_{\text{en}})), \\
\mathcal{S}^{3}_{\text{en}} =&\ {\small \text{\sffamily{3DSwinTrans}}}({\small{\text{\sffamily{PatchMer}}}}(\mathcal{S}^{2}_{\text{en}})), 
\end{aligned}
\right.
\end{equation}
where $\small{\text{\sffamily{3DPatchPar}}}(\cdot)$ and $\small{\text{\sffamily{LinearEm}}}(\cdot)$ are a 3D patch partition and a linear embedding layer (see Section \ref{4.2}), respectively, $\small{\text{\sffamily{3DSwinTrans}}}(\cdot)$ is the consecutive 3D Swin Transformer blocks (see Section \ref{4.4}), and $\small{\text{\sffamily{PatchMer}}}(\cdot)$ is the patch merging layer (see Section \ref{4.3}). Further, $\mathcal{S}^{i}_{\text{en}}$, $i \in \{1, 2, 3\}$ stands for the extracted spatiotemporal usage pattern features after $i$th $\small{\text{\sffamily{3DSwinTrans}}}(\cdot)$ in $S_{\bm{\alpha}^{\text{3D}}}$, $\mathcal{X}_{\text{en}}$ are the non-overlapping 3D patches, and $\mathcal{S}^{3}_{\text{en}}$ is also the output of $S_{\bm{\alpha}^{\text{3D}}}$. Then, there is a bottleneck layer $\small{\text{\sffamily{bottleneck}}}(\cdot)$ between $S_{\bm{\alpha}^{\text{3D}}}$ and $D^{\text{3D}}_{\bm{\beta}^{\text{3D}}}$, which consists of the same number of 3D Swin Transformer blocks as the third layer of $S_{\bm{\alpha}^{\text{3D}}}$. The process of features passing through a bottleneck layer is $\mathcal{X}_{\text{de}} = \ {\small \text{\sffamily{bottleneck}}}(\mathcal{S}^{3}_{\text{en}})$. 

\textit{Predictor $D^{\text{3D}}_{\bm{\beta}^{\text{3D}}}$:} As illustrated in Fig. \ref{Fig:3D-SwinM}, $D^{\text{3D}}_{\bm{\beta}^{\text{3D}}}$ maintains a symmetrical structure with $S_{\bm{\alpha}^{\text{3D}}}$. Note that the patch merging layer is replaced by the patch expanding layer, and a 3D projection layer is followed by the third layer of $D^{\text{3D}}_{\bm{\beta}^{\text{3D}}}$, which maps the decoding features to the future spectrograms. Compared with the down-sampling of the patch merging layer, the patch expanding layer performs the up-sampling operation. Moreover, the extracted features are fused with multi-scale features from $S_{\bm{\alpha}^{\text{3D}}}$ via multiple skip connections. The process of predicting can be given by      
\begin{equation}\label{eq6}
\left\{
\begin{aligned}
\mathcal{S}^{1}_{\text{Tr}}=&\ \small{\text{\sffamily{3DSwinTrans}}}(\small{\text{\sffamily{Concat}}}(\mathcal{X}_{\text{de}},\mathcal{S}^{3}_{\text{en}})), \\
\mathcal{S}^{1}_{\text{de}}=&\ \small{\text{\sffamily{Concat}}}(\small{\text{\sffamily{PatchExp}}}(\mathcal{S}^{1}_{\text{Tr}}), \mathcal{S}^{2}_{\text{en}}), \\
\mathcal{S}^{2}_{\text{de}}=&\ \small{\text{\sffamily{Concat}}}(\small{\text{\sffamily{PatchExp}}}(\small{\text{\sffamily{3DSwinTrans}}}(\mathcal{S}^{1}_{\text{de}})), \mathcal{S}^{1}_{\text{en}}), \\
\mathcal{S}^{3}_{\text{de}}=&\ \small{\text{\sffamily{3DProjectLayer}}}(\small{\text{\sffamily{3DSwinTrans}}}(\mathcal{S}^{2}_{\text{de}})),  
\end{aligned}
\right.
\end{equation}
where $\small{\text{\sffamily{PatchExp}}}(\cdot)$ is the patch expanding layer (see Section \ref{4.3}), $\small{\text{\sffamily{Concat}}}(\cdot, \cdot)$ is to concatenate two tensors in $C$ (i.e., skip connection), and $\small{\text{\sffamily{3DProjectLayer}}}(\cdot)$ is the 3D projection layer (see Section \ref{4.5}). Further, $\mathcal{S}^{1}_{\text{Tr}}$ is the mapping features of the first $\small{\text{\sffamily{3DSwinTrans}}}(\cdot)$ in $D^{\text{3D}}_{\bm{\beta}^{\text{3D}}}$, $\mathcal{S}^{n}_{\text{de}}$, $n \in \{1, 2, 3\}$ stands for the decoded features after the $n$th decoding layer in $D^{\text{3D}}_{\bm{\beta}^{\text{3D}}}$, and $\mathcal{S}^{3}_{\text{de}}$ is also the final output. The predicted spectrograms $\mathcal{S}^{3}_{\text{de}}$ ($\hat{\bm{{\rm X}}}^{*}_{T+1:T+K}$) are used in a spectrum monitoring task within a spectrum management entity to detect possible anomalies. 

From (\ref{eq5}) and (\ref{eq6}), flow processing strategy refers to the symmetric spectrum data processing flow from feature extraction to feature prediction, constructed using 3D-ViT blocks as components. The two unique designs of this process are the 3D-ViT blocks and the symmetric designs. Here, the symmetric designs are: encoder's structure $\{2, 4, 2\}$ to predictor's structure $\{2, 4, 2\}$ and patch merging to patch expanding. 

\textit{Pyramid:} From the description of $S_{\bm{\alpha}^{\text{3D}}}$ and $D^{\text{3D}}_{\bm{\beta}^{\text{3D}}}$, the 3D-SwinSTB employs two key components, bottleneck layer, and skip connection, within the feature pyramid to assist in inferring future spectrograms using a flow processing strategy. Skip connections provide direct pathways for gradients to flow from deeper layers to the shallower layers, allow the high-resolution spectrum features from encoder to be directly transferred to predictor, and help model learn both low-level details and high-level abstractions simultaneously. These functions can reduce the risk of losing critical features during the process of layer-by-layer propagation. The bottleneck layer can capture and consolidate high-level spectrum usage dependencies and reduce the computational burden by compressing features.

\subsection{3D Patch Partition and Linear Embedding Layer}\label{4.2}
To handle $T$-frame spectrograms, we reshape the $T$-frame spectrograms $ \bm{{\rm X}}_{1:T} \in \mathbb{R}^{T \times H \times W \times 3}$ into a sequence of flattened 3D patches $\mathcal{X}_{\text{en}} \in \mathbb{R}^{N \times (3T_p H_p W_p)}$. Here, $(T_p, H_p, W_p)$ is the resolution of each 3D patch, $N=THW/T_p H_p W_p$ is the number of the 3D patches, and each 3D patch comprises a $3T_p H_p W_p$-dimensional feature. The above process is summarized as the 3D patch partition. For the linear embedding layer, we follow the configuration in \cite{dosovitskiy2020image}. Specifically, a 3D convolution layer is applied to project the features of each 3D patch to an arbitrary dimension denoted by $C$, and these 3D patches are then fed into a 3D Swin Transformer layer.   

\subsection{Patch Merging Layer and Patch Expanding Layer}\label{4.3}
\textit{Patch merging layer:} We use the input patches $\mathcal{S}^{1}_{\text{en}} \in \mathbb{R}^{\frac{T}{T_p} \times \frac{H}{H_p} \times \frac{W}{W_p} \times C}$ as an example to introduce this layer. This layer first concatenates the features of each group of $2 \times 2$ spatially neighboring 3D patches in $\mathcal{S}^{1}_{\text{en}}$. This reduces the number of 3D tokens by a multiple of $2 \times 2 = 4$, i.e., $2 \times$ down-sampling of resolution. Then this layer applies a simple linear layer $f_l$ to project the 4$C$-dimensional concatenated features $\mathcal{S}^{1,\text{con}}_{\text{en}} \in \mathbb{R}^{\frac{T}{T_p} \times \frac{H}{2H_p} \times \frac{W}{2W_p} \times 4C}$ to dimension 2$C$:  
\begin{equation}\label{Eq11}
\hat{\mathcal{S}}^{1,\text{con}}_{\text{en}} = f_l(\mathcal{S}^{1,\text{con}}_{\text{en}}), 
\end{equation}
where $\hat{\mathcal{S}}^{1,\text{con}}_{\text{en}} \in \mathbb{R}^{\frac{T}{T_p} \times \frac{H}{2H_p} \times \frac{W}{2W_p} \times 2C}$. This layer aggregates contextual features by merging patches from different parts of the spectrogram, reduces computational complexity by lowering the spatial resolution, and enhances feature representation by increasing $C$. These operations enable the 3D Swin Transformer block to better learn the relationships between spectrum usage patterns in different time-frequency parts.  

\textit{Patch expanding layer:} This layer performs the opposite of the patch merging layer. Taking the first patch expanding layer as an example, a linear layer $f_l$ is first applied on the input features $\mathcal{S}^{\text{exp}}_{\text{de}}  \in \mathbb{R}^{\frac{T}{T_p} \times \frac{H}{4H_p} \times \frac{W}{4W_p} \times 4C}$ (i.e., $\mathcal{S}^{1}_{\text{Tr}}$):
\begin{equation}\label{Eq111}
\hat{\mathcal{S}}^{\text{exp}}_{\text{de}} = f_l(\mathcal{S}^{\text{exp}}_{\text{de}}), 
\end{equation}
where $\hat{\mathcal{S}}^{\text{exp}}_{\text{de}} \in \mathbb{R}^{\frac{T}{T_p} \times \frac{H}{4H_p} \times \frac{W}{4W_p} \times 8C}$. From (\ref{Eq111}), the feature dimension is increased to $2 \times$ by the original dimension. We then rearrange $\hat{\mathcal{S}}^{\text{exp}}_{\text{de}}$ to extend its resolution by a factor of 2 while reducing the feature dimension to a quarter of its original size using a \textit{rearrange} function in Pytorch, and the resolution and feature dimension of $\hat{\mathcal{S}}^{\text{exp}}_{\text{de}}$ become $\frac{T}{T_p} \times \frac{H}{2H_p} \times \frac{W}{2W_p} \times 2C$. After the above process, the size of input $\mathcal{S}^{\text{exp}}_{\text{de}}$ is changed from $\frac{T}{T_p} \times \frac{H}{4H_p} \times \frac{W}{4W_p} \times 4C$ to $\frac{T}{T_p} \times \frac{H}{2H_p} \times \frac{W}{2W_p} \times 2C$, achieving a $2\times$ up-sampling resolution. This layer refines the features extracted from low-resolution spectrogram and increases spatial resolution. These operations help the 3D Swin Transformer to infer details and spatial relationships of usage patterns in future spectrograms.

\subsection{3D Swin Transformer Block}\label{4.4}
\begin{figure}
	\centerline{\includegraphics[width=88mm,height=28mm]{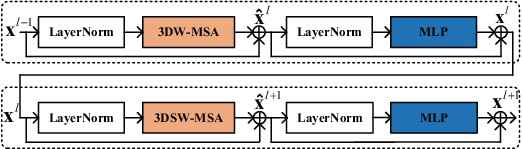}}
	\caption{The illustration of two successive 3D Swin Transformer blocks.}
	\label{Fig:Swin}
\end{figure}
As illustrated in Fig. \ref{Fig:Swin}, two successive 3D Swin Transformer blocks include a standard 3D multi-head self-attention (3D-MSA) module with non-overlapping 3D window (denoted as $\small{\text{\sffamily{3DW-MSA}}}(\cdot)$), a MSA module with non-overlapping 3D shifted window (denoted as $\small{\text{\sffamily{3DSW-MSA}}}(\cdot)$) \cite{liu2021swin}, and a feed-forward network (FFN, denoted as ${\small\text{\sffamily{MLP}}}(\cdot)$, is a 2-layer multilayer perceptron (MLP), with Gaussian error linear unit (GELU) \cite{chen2021remote} non-linearity in between) following each MSA module. Layer normalization (LN, denoted as ${\small \text{\sffamily{LN}}}(\cdot)$) is applied before each MSA module and FFN, and a residual connection is applied after each module. The implementation process is      
\begin{equation}\label{eq7}
\left\{
\begin{aligned}
\hat{\bm{{\rm x}}}^{l}=&\ \small{\text{\sffamily{3DW-MSA}}}(\text{\sffamily{LN}}(\bm{{\rm x}}^{l-1}))+\bm{{\rm x}}^{l-1}, \\
\bm{{\rm x}}^{l}=&\ \small{\text{\sffamily{MLP}}}(\text{\sffamily{LN}}(\hat{\bm{{\rm x}}}^{l})) + \hat{\bm{{\rm x}}}^{l}, \\
\hat{\bm{{\rm x}}}^{l+1}=&\ \small{\text{\sffamily{3DSW-MSA}}}(\text{\sffamily{LN}}(\bm{{\rm x}}^{l}))+\bm{{\rm x}}^{l}, \\
\bm{{\rm x}}^{l+1}=&\ \small{\text{\sffamily{MLP}}}(\text{\sffamily{LN}}(\hat{\bm{{\rm x}}}^{l+1}))+\hat{\bm{{\rm x}}}^{l+1}, 
\end{aligned}
\right.
\end{equation}
where $\hat{\bm{{\rm x}}}^{l}$ and $\bm{{\rm x}}^{l}$ are the output features of the $\small{\text{\sffamily{3DW-MSA}}}(\cdot)$ module and the FFN for $l$th block, respectively, $\hat{\bm{{\rm x}}}^{l+1}$ and $\bm{{\rm x}}^{l+1}$ are the output features of the $\small{\text{\sffamily{3DSW-MSA}}}(\cdot)$ module and the FFN for $(l+1)$th block, respectively,
\begin{equation}\label{Eq8}
\small{\text{\sffamily{MLP}}}(\text{\sffamily{LN}}(\hat{\bm{{\rm x}}}^{l})) = \text{GELU}(\text{\sffamily{LN}}(\hat{\bm{{\rm x}}}^{l})\bm{{\rm W}}_1+\bm{{\rm b}}_1)\bm{{\rm W}}_2+\bm{{\rm b}}_2,
\end{equation}
$\bm{{\rm W}}_1 \in \mathbb{R}^{C \times 2C}$ and $\bm{{\rm W}}_2 \in \mathbb{R}^{2C \times C}$ stand for the linear projection matrices with biases $\bm{{\rm b}}_1$ and $\bm{{\rm b}}_2$. $\small{\text{\sffamily{3DW-MSA}}}(\cdot)$ and $\small{\text{\sffamily{3DSW-MSA}}}(\cdot)$ are the two core components of the 3D Swin Transformer block. While both calculate self-attention in non-overlapping 3D windows, the $\small{\text{\sffamily{3DSW-MSA}}}(\cdot)$ uses a \textit{shifted} window. Next, we use an example to illustrate the difference between a normal window and a shifted window. For the former, given a spectrogram video composed of $T \times H \times W$ 3D tokens and a 3D window size of $P \times M \times M$\footnote{To make the window size $(P, M, M)$ divisible by the feature map size of $(T, H, W)$, bottom-right padding is employed on the feature map if needed.}. The video tokens are partitioned into $\lceil \frac{T}{P} \rceil \times \lceil \frac{H}{M} \rceil \times \lceil \frac{W}{M} \rceil$ non-overlapping 3D windows. We assume that the input size is $16 \times 16 \times 16$ and the window size is $8 \times 8 \times 8$, the number of windows in layer $l$ would be $8$. For the latter, given the same input size and window size as the former, the self-attention module's window partition configuration in the $(l+1)$th layer is shifted along the time, height and width axes by $(\frac{P}{2},\frac{M}{2},\frac{M}{2})$ tokens from that of the $l$th layer’s self-attention module. So, the number of windows becomes $3 \times 3 \times 3 = 27$. To address the increased number of windows, we use the efficient batch computation\footnote{It is implemented by the cyclic-shift and masking mechanism, which can be found in \cite{liu2021swin}.} to make the number of 3D shifted windows the same as the number of 3D windows.

From  (\ref{eq7}), the core designs of 3D-SwinSTB that capture local and global spectrum usage patterns are $\small{\text{\sffamily{3DW-MSA}}}(\cdot)$ function and $\small{\text{\sffamily{3DSW-MSA}}}(\cdot)$ function, respectively. Firstly, the $\small{\text{\sffamily{3DW-MSA}}}(\cdot)$ function is
\begin{equation}\label{Eq2}
{\small{\text{\sffamily{3DW-MSA}}}}({\small\text{\sffamily{LN}}}(\bm{{\rm x}}^{l-1})) = \text{Concat}(\text{head}^{l}_1, \cdots ,\text{head}^{l}_{\zeta})\bm{{\rm W}}^l.
\end{equation}
Here,
\begin{equation}\label{Eq3}
\begin{aligned}
\text{head}^{l}_i=& {\small{\text{\sffamily{3DW-Attention}}}}(\bm{{\rm Q}}^{l-1}_i,\bm{{\rm K}}^{l-1}_i,\bm{{\rm V}}^{l-1}_i)\\
=& \sigma (\frac{\bm{{\rm Q}}^{l-1}_i{\bm{{\rm K}}^{l-1}_i}^{T}}{\sqrt d} + \bm{{\rm B}})\bm{{\rm V}}^{l-1}_i,
\end{aligned}
\end{equation}
where $i \in \{1, \cdots, \zeta\}$, $\bm{{\rm Q}}^{l-1}_i = \text{\sffamily{LN}}(\bm{{\rm x}}^{l-1}){\bm{{\rm W}}^{l-1}_{i}}^{(q)}$ (\textit{query}), $\bm{{\rm K}}^{l-1}_i = \text{\sffamily{LN}}(\bm{{\rm x}}^{l-1}){\bm{{\rm W}}^{l-1}_{i}}^{(k)}$ (\textit{key}), $\bm{{\rm V}}^{l-1}_i = \text{\sffamily{LN}}(\bm{{\rm x}}^{l-1}){\bm{{\rm W}}^{l-1}_{i}}^{(v)}$ (\textit{value}) $\in \mathbb{R}^{PM^2 \times d}$, and $\bm{{\rm B}} \in \mathbb{R}^{ P^2 \times M^2 \times M^2}$ is a 3D relative position bias\footnote{Since the relative position along each axis lies in the range of $[-P+1, P-1]$ (in time) or $[-M+1, M-1]$ (in height or width), we parameterize a smaller-sized bias matrix $\hat{\bm{{\rm B}}} \in \mathbb{R}^{(2T-1)\times(2M-1)\times(2M-1)}$, and values in $\bm{{\rm B}}$ are taken from $\hat{\bm{{\rm B}}}$ \cite{hu2019local}.}. Here, $P \times M^2$ is the number of patches in a 3D window, ${\bm{{\rm W}}^{l-1}_{i}}^{(q)}$, ${\bm{{\rm W}}^{l-1}_{i}}^{(k)}$, ${\bm{{\rm W}}^{l-1}_{i}}^{(v)} \in \mathbb{R}^{C \times d}$ are the learnable parameters of three linear projection layers for $l$-th block, $d$ is the dimension of \textit{query}/\textit{key}, $\bm{{\rm W}}^l \in \mathbb{R}^{\zeta d \times C}$ are the linear projection matrices for $l$-th block, $C$ is the number of channels, $\zeta$ is the number of heads, and $\sigma$ is the softmax function. Secondly, the ${\small \text{\sffamily{3DSW-MSA}}}(\cdot)$ function is
\begin{equation}\label{Eq5}
{\small{\text{\sffamily{3DSW-MSA}}}}({\small{\text{\sffamily{LN}}}}(\bm{{\rm x}}^{l})) = \text{Concat}(\text{head}^{l+1}_1, \cdots ,\text{head}^{l+1}_{\zeta})\bm{{\rm W}}^{l+1}.
\end{equation}
Here, 
\begin{equation}\label{Eq6}
\begin{aligned}
\text{head}^{l+1}_i=& {\small{\text{\sffamily{3DSW-Attention}}}}(\bm{{\rm Q}}^l_i,\bm{{\rm K}}^l_i,\bm{{\rm V}}^l_i)\\
=& \sigma (\frac{\bm{{\rm Q}}^l_i{\bm{{\rm K}}^l_i}^{T}}{\sqrt d} + \bm{{\rm B}})\bm{{\rm V}}^l_i,
\end{aligned}
\end{equation}
where $i \in \{1, \cdots, \zeta\}$, $\bm{{\rm Q}}^l_i = \text{\sffamily{LN}}(\bm{{\rm x}}^{l}){\bm{{\rm W}}^{l}_{i}}^{(q)}$ (\textit{query}), $\bm{{\rm K}}^l_i = \text{\sffamily{LN}}(\bm{{\rm x}}^{l}){\bm{{\rm W}}^{l}_{i}}^{(k)}$ (\textit{key}), $\bm{{\rm V}}^l_i = \text{\sffamily{LN}}(\bm{{\rm x}}^{l}){\bm{{\rm W}}^{l}_{i}}^{(v)}$ (\textit{value}) $\in \mathbb{R}^{PM^2 \times d}$. Here, ${\bm{{\rm W}}^{l}_{i}}^{(q)}$, ${\bm{{\rm W}}^{l}_{i}}^{(k)}$, ${\bm{{\rm W}}^{l}_{i}}^{(v)} \in \mathbb{R}^{C \times d}$ are the learnable parameters of three linear projection layers and $\bm{{\rm W}}^{l+1} \in \mathbb{R}^{\zeta d \times C}$ are the linear projection matrices for $(l+1)$-th block. Different from (\ref{Eq2}), self-attention is calculated again by (\ref{Eq5}) in the 3D shifted windows. This process is similar to calculating self-attention in the original 3D windows, but the window positions have been changed.

The 3D Swin Transformer block is a key component in the proposed symmetric flow processing strategy, which adopts a 3D window, a 3D shifted window and a 3D-MSA designs. These designs come with the following advantages:

\textit{Advantage 1:} Supposing each 3D window contains $P \times M \times M$ 3D patches, the computational complexity of a global (standard) 3D MSA module and a 3D window based on a spectrogram video of $p\times h \times w$ 3D patches are \cite{liu2021swin}
\begin{equation}\label{EqMSA}
\mathcal{O}(\text{3D-MSA}) = 4phwC^2 + 2(phw)^2C,
\end{equation}
\begin{equation}\label{EqWSMA}
\mathcal{O}(\text{3DW-MSA}) = 4phwC^2 + 2PM^2phwC,
\end{equation}
where the 3D-MSA is quadratic to 3D patch number $phw$, and 3DW-MSA is linear when $P$ and $M$ are fixed (set to 2 and 7 by default, respectively). From (\ref{EqMSA}) and (\ref{EqWSMA}), the 3D window can improve the efficiency of the model for a large $phw$.

\textit{Advantage 2:} The 3DW-MSA operates within individual 3D windows, but lacks inter-window connectivity, potentially limiting the model's representational capacity. In contrast, the 3D shifted window design establishes cross-window connections while preserving computational efficiency. This allows the model to learn usage patterns across both local and global frequency bands in the spectrogram.

\textit{Advantage 3:} The 3D-MSA design can continuously learn the spatiotemporal proximity and contextual trends of usage behavior between different frequency bands in a multi-frame spectrogram series, which provides accurate detailed changes for spectrum monitoring tasks. However, traditional ViT only gives different attention in a single spectrogram, and cannot capture the long-term dependence of user behavior.

\subsection{3D Projection Layer}\label{4.5}
The 3D projection layer uses multiple 3D inverse convolutional layers to project the mapping features (denoted as $\mathcal{S}^{3}_{\text{Tr}}$) of third $\small{\text{\sffamily{3DSwinTrans}}}(\cdot)$ in $D^{\text{3D}}_{\bm{\beta}^{\text{3D}}}$ into the future spectrograms $\hat{\bm{{\rm X}}^{*}}_{T+1:T+K}$. The specific process can be given by
\begin{equation}\label{Eq12}
\begin{aligned}
\mathcal{S}^{3}_{\text{Tr}} \xrightarrow[(t_k,h_k,w_k), (t_s,h_s,w_s)]{\text{3D Conv}^{-1}} {\mathcal{S}^{3}_{\text{Tr}}}^{'} 
\text{Re}^{n}(\xrightarrow[(1,1,1), (1,1,1)]{\text{3D Conv}^{-1}}) \hat{\bm{{\rm X}}}^{*}_{T+1:T+K},
\end{aligned}
\end{equation}
the resolution and feature dimension changes of $\mathcal{S}^{3}_{\text{Tr}}$ are
\begin{equation}\label{Eq122}
\frac{T}{T_p} \times \frac{H}{H_p} \times \frac{W}{W_p} \times C \rightarrow T \times H \times W \times C \rightarrow 
T \times H \times W \times 3,
\end{equation}
where $\xrightarrow[(t_k,h_k,w_k), (t_s,h_s,w_s)]{\text{3D Conv}^{-1}}$ stands for the 3D inverse convolution operation with convolution kernel size of $(t_k,h_k,w_k)$ and stride size of $(t_s,h_s,w_s)$. Herein, $(t_k,h_k,w_k) = (t_s,h_s,w_s) = (T_p, H_p, W_p)$. Furthermore, $\text{Re}^{n}(\cdot)$ stands for the operation is repeated $n$ ($n=\lceil\text{log}_2(\frac{C}{3})\rceil$) times. The convolution kernel and stride sizes of the 3D inverse convolution operation in $\text{Re}^{n}(\cdot)$ default to 1.
\begin{algorithm}[t]
	\caption{Train 3D-SwinSTB Algorithm} 
	\label{alg:3D-SwinSTB}
	\hspace*{0.04in}{\textbf{Initialization}: Initial parameters $\{\bm{\alpha}^{\text{3D}}, \bm{\beta}^{\text{3D}}, \bm{{\rm b}}\}$.}
	
	\begin{algorithmic}[1] 
		\STATE \textbf{Input}: The historical spectrograms $\bm{{\rm X}}_{1:T}$;
		\STATE \textbf{while} $\mathcal{L}$ decreases by less than $\vartheta_{\text{per}}$ \% in $n_{\text{ep}}$ epochs \textbf{do}
		\STATE \quad $\mathcal{X}_{\text{en}} =\ \small{\text{\sffamily{3DPatchPar}}}(\bm{{\rm X}}_{1:T})$;\\
		\STATE \quad $\mathcal{S}^{1}_{\text{en}} =\ \small{\text{\sffamily{3DSwinTrans}}}(\small{\text{\sffamily{LinearEm}}}(\mathcal{X}_{\text{en}}))$;\\
		\STATE \quad $\mathcal{S}^{2}_{\text{en}} =\ \small{\text{\sffamily{3DSwinTrans}}}(\small{\text{\sffamily{PatchMer}}}(\mathcal{S}^{1}_{\text{en}}))$; \\
		\STATE \quad $\mathcal{S}^{3}_{\text{en}} =\ \small{\text{\sffamily{3DSwinTrans}}}(\small{\text{\sffamily{PatchMer}}}(\mathcal{S}^{2}_{\text{en}}))$; \\
		\STATE \quad // Encoder $S_{\bm{\alpha}^{\text{3D}}}$;\\ 
		\STATE \quad $\mathcal{X}_{\text{de}} = \ \small{\text{\sffamily{bottleneck}}}(\mathcal{S}^{3}_{\text{en}})$;\\
		\STATE \quad // Bottleneck layer;\\
		\STATE \quad $\mathcal{S}^{1}_{\text{Tr}}=\ \small{\text{\sffamily{3DSwinTrans}}}(\small{\text{\sffamily{Concat}}}(\mathcal{X}_{\text{de}},\mathcal{S}^{3}_{\text{en}}))$; \\
		\STATE \quad $\mathcal{S}^{1}_{\text{de}}=\ \small{\text{\sffamily{Concat}}}(\small{\text{\sffamily{PatchExp}}}(\mathcal{S}^{1}_{\text{Tr}}), \mathcal{S}^{2}_{\text{en}})$; \\
		\STATE \quad $\mathcal{S}^{2}_{\text{de}}=\ \small{\text{\sffamily{Concat}}}(\small{\text{\sffamily{PatchExp}}}(\small{\text{\sffamily{3DSwinTrans}}}(\mathcal{S}^{1}_{\text{de}})), \mathcal{S}^{1}_{\text{en}})$; \\
		\STATE \quad $\mathcal{S}^{3}_{\text{de}}=\ \small{\text{\sffamily{3DProjectLayer}}}(\small{\text{\sffamily{3DSwinTrans}}}(\mathcal{S}^{2}_{\text{de}}))$;\\ 
		\STATE \quad // Predictor $D^{\text{3D}}_{\bm{\beta}^{\text{3D}}}$;\\
		\STATE \quad Compute loss function $\mathcal{L}$ by (\ref{loss});\\
		\STATE \quad Train $\{\bm{\alpha}^{\text{3D}}, \bm{\beta}^{\text{3D}}, \bm{{\rm b}}\}$ by optimizer with (\ref{eq25});\\
		\STATE \textbf{end while}  
		\STATE Obtain updated parameters $\{{{\bm{\alpha}}^{*}}^{\text{3D}}, {{\bm{\beta}}^{*}}^{\text{3D}}, {\bm{{\rm b}}}^{*}\}$;\\ 
		\STATE \textbf{Output}: The trained 3D-SwinSTB ${D}^{\text{3D}}_{{\bm{\beta}^{*}}^{\text{3D}}}(S_{{\bm{\alpha}^{*}}^{\text{3D}}}(\cdot))$.	
	\end{algorithmic}
\end{algorithm}

\subsection{Training Rules}\label{4.6}
Before spectrogram prediction, we need to train the proposed 3D-SwinSTB with rules. The aim of training the 3D-SwinSTB is minimize the error between $\bm{{\rm X}}^{*}_{T+1:T+K}$ and $\hat{\bm{{\rm X}}}^{*}_{T+1:T+K}$ by continuously updating the values of $\bm{\alpha}^{\text{3D}}$, $\bm{\beta}^{\text{3D}}$, and $\bm{{\rm b}}$ ($\bm{{\rm b}}$ is the bias of the 3D-SwinSTB). Here, the learnable parameters of the bottleneck layer are included in $\bm{\alpha}$ for the convenience of analysis. The mean squared error (MSE) is used as the error function to measure the difference between $\bm{{\rm X}}^{*}_{T+1:T+K}$ and $\hat{\bm{{\rm X}}}^{*}_{T+1:T+K}$, which can be formulated as
\begin{equation}\label{loss}
\mathcal{L}(\bm{{\rm X}}^{*}_{T+1:T+K},\hat{\bm{{\rm X}}}^{*}_{T+1:T+K}) = \parallel \bm{{\rm X}}^{*}_{T+1:T+K} - \hat{\bm{{\rm X}}}^{*}_{T+1:T+K} \parallel^{2}_{2}.
\end{equation}

Let $\bm{{\rm U}} = \{\bm{\alpha}^{\text{3D}}, \bm{\beta}^{\text{3D}}, \bm{{\rm b}}\}$, the optimal training parameters $\bm{{\rm U}}^{*} = \{{{\bm{\alpha}}^{*}}^{\text{3D}}, {{\bm{\beta}}^{*}}^{\text{3D}}, {\bm{{\rm b}}}^{*}\}$ of 3D-SwinSTB can be obtained by
\begin{equation}
\bm{{\rm U}}^{*} = {\rm arg} \mathop {\rm min} \limits_{\bm{{\rm U}}} \mathcal{L}(\bm{{\rm X}}^{*}_{T+1:T+K},\hat{\bm{{\rm X}}}^{*}_{T+1:T+K}).
\end{equation}
Specifically, we use a gradient-based optimizer (follow Sec. \ref{setup}) to get the optimal $\bm{{\rm U}}^{*}$, which can be given by
\begin{equation}\begin{split}\label{eq25}
\bm{\alpha}^{\text{3D}}(s+1) \leftarrow& \bm{\alpha}^{\text{3D}}(s)-\mu \dfrac{\partial \mathcal{L}(\bm{{\rm X}}^{*}_{T+1:T+K},\hat{\bm{{\rm X}}}^{*}_{T+1:T+K})}{\partial \bm{\alpha}^{\text{3D}}},\\
\bm{\beta}^{\text{3D}}(s+1) \leftarrow& \bm{\beta}^{\text{3D}}(s)-\mu \dfrac{\partial \mathcal{L}(\bm{{\rm X}}^{*}_{T+1:T+K},\hat{\bm{{\rm X}}}^{*}_{T+1:T+K})}{\partial \bm{\beta}^{\text{3D}}},\\
\bm{{\rm b}}(s+1) \leftarrow& \bm{{\rm b}}(s)-\mu \dfrac{\partial \mathcal{L}(\bm{{\rm X}}^{*}_{T+1:T+K},\hat{\bm{{\rm X}}}^{*}_{T+1:T+K})}{\partial \bm{{\rm b}}},
\end{split}\end{equation}
where $\bm{\alpha}^{\text{3D}}(s)$, $\bm{\beta}^{\text{3D}}(s)$, and $\bm{{\rm b}}(s)$ are weights and biases of the $s$th training step, and $\mu$ is the learning rate. The training process of the 3D-SwinSTB can be found in Algorithm \ref{alg:3D-SwinSTB}.
\begin{figure}
	\centerline{\includegraphics[width=90mm,height=38mm]{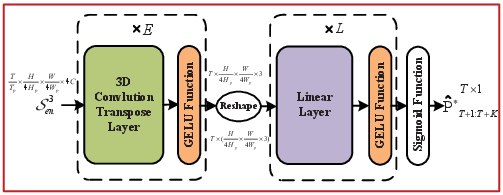}}
	\caption{The structure of a simple SOR predictor $D^{\text{SOR}}_{\bm{\beta}^{\text{SOR}}}$.}
	\label{Fig:Linear}
\end{figure}

\section{Proposed 3D-SwinLinear Method Design \\ for SOR prediction}\label{sec5}
\subsection{Definition of SOR}\label{5.1}
We first give the SOR definition. Given the occupancy status (idle 0 or occupied 1) of $F$ frequency bands at time $t$, the frequency band occupancy rate can be represented as $P_F = 1-F^{(0)}/F$, where $F^{(0)}$ is the number of idle channels; given the occupancy status of $T$ sampling timeslots at frequency$f$, the time occupancy rate can be represented as $P_T = 1-T^{(0)}/T$, where $T^{(0)}$ is the number of idle timeslots; given the occupancy status of $F$ frequency channels with $T$ time slots, the SOR can be represented as $P_{\text{sor}} = 1-F^{(0)}T^{(0)}/FT$ \cite{6933906}.

In this paper, we use supervised learning to predict the future SOR based on the spectrograms. Therefore, according to the SOR definition, the SOR is estimated by the statistic of gray binarization of the spectrogram \cite{7120120}. The local average method is used to calculate the threshold of binarization. Firstly, the spectrogram is divided into several small blocks, and then the average value of pixels is calculated in each small block. Finally, the global threshold is calculated according to the average value. The local threshold is calculated by
\begin{equation}
\theta(x,y) = \dfrac{1}{w^2} \sum_{i=x-w/2}^{x+w/2} \sum_{j=y-w/2}^{y+w/2}I(i,j),
\end{equation}
where $I(x,y)$ represents the pixel value of the image at position $(x,y)$ and $w$ is the size of the small block.
\begin{algorithm}
	\caption{Train 3D-SwinLinear Algorithm} 
	\label{alg:3D-SwinLinear}
	\hspace*{0.04in}{\textbf{Initialization}: Initial parameter set $\{\bm{\alpha}^{\text{SOR}}, \bm{\beta}^{\text{SOR}}, \bm{{\rm b}}\}$.}
	
	\begin{algorithmic}[1] 
		\STATE \textbf{Input}: The historical spectrograms $\bm{{\rm X}}_{1:T}$;
		
		\STATE \textbf{while} $\mathcal{L}$ decreases by less than $\vartheta_{\text{per}}$ \% in $n_{\text{ep}}$ epochs \textbf{do}
		\STATE \quad $\mathcal{X}_{\text{en}} = \ \small{\text{\sffamily{3DPatchPar}}}(\bm{{\rm X}}_{1:T})$; \\
		\STATE \quad $\mathcal{S}^{1}_{\text{en}} = \ \small{\text{\sffamily{3DSwinTrans}}}(\small{\text{\sffamily{LinearEm}}}(\mathcal{X}_{\text{en}}))$; \\
		\STATE \quad $\mathcal{S}^{2}_{\text{en}} = \ \small{\text{\sffamily{3DSwinTrans}}}(\small{\text{\sffamily{PatchMer}}}(\mathcal{S}^{1}_{\text{en}}))$; \\
		\STATE \quad $\mathcal{S}^{3}_{\text{en}} = \ \small{\text{\sffamily{3DSwinTrans}}}(\small{\text{\sffamily{PatchMer}}}(\mathcal{S}^{2}_{\text{en}}))$; \\
		\STATE \quad // Encoder $S_{\bm{\alpha}^{\text{SOR}}}$;
		\STATE \quad $\hat{\mathcal{S}}^{1}_{\text{de}}=\ \small{\text{\sffamily{3DConvTransGELU}}}^{(E)}(\mathcal{S}^{3}_{\text{en}})$; \\
		\STATE \quad $\hat{\mathcal{S}}^{2}_{\text{de}}=\ \small{\text{\sffamily{LinearGELU}}}^{(L)}(\hat{\mathcal{S}}^{1}_{\text{de}})$; \\
		\STATE \quad $\hat{{\rm P}}^{*}_{T+1:T+K}=\ \small{\text{\sffamily{Sigmoid}}}(\hat{\mathcal{S}}^{2}_{\text{de}})$;  
		\STATE \quad // Predictor $D^{\text{SOR}}_{\bm{\beta}^{\text{SOR}}}$;\\
		\STATE \quad Compute loss function $\mathcal{L}$ by (\ref{loss});\\
		\STATE \quad Train $\bm{\alpha}^{\text{SOR}}, \bm{\beta}^{\text{SOR}}, \bm{{\rm b}}$ by optimizer with (\ref{eq25});\\
		\STATE \textbf{end while}
		\STATE Obtain updated parameters $\{{{\bm{\alpha}}^{*}}^{\text{SOR}}, {{\bm{\beta}}^{*}}^{\text{SOR}}, {\bm{{\rm b}}}^{*}\}$;\\ 
		\STATE \textbf{Output}: The trained 3D-SwinLinear ${D}^{\text{SOR}}_{{\bm{\beta}^{*}}^{\text{SOR}}}(S_{{\bm{\alpha}^{*}}^{\text{SOR}}}(\cdot))$.	
	\end{algorithmic}
\end{algorithm}

\subsection{Design of Algorithm}\label{5.2}
To solve (\ref{Eq_ssp}), 3D-SwinLinear is designed, which consists of a 3D-SwinSTB's encoder $S_{\bm{\alpha}^{\text{SOR}}}$ with a parameter set $\bm{\alpha}^{\text{SOR}}$ and a SOR predictor $D^{\text{SOR}}_{\bm{\beta}^{\text{SOR}}}$ with a simple structure composed of 3D convolutions and linear layers. As shown in Fig. \ref{Fig:Linear}, the specific implementation process of $D^{\text{SOR}}_{\bm{\beta}^{\text{SOR}}}$ can be given by 
\begin{equation}\label{eq15}
\left\{
\begin{aligned}
&\hat{\mathcal{S}}^{1}_{\text{de}}=\ \small{\text{\sffamily{3DConvTransGELU}}}^{(E)}(\mathcal{S}^{3}_{\text{en}}), \\
&\hat{\mathcal{S}}^{2}_{\text{de}}=\ \small{\text{\sffamily{LinearGELU}}}^{(L)}(\small{\text{\sffamily{Reshape}}}(\hat{\mathcal{S}}^{1}_{\text{de}})), \\
&\hat{{\rm P}}^{*}_{T+1:T+K}=\ \small{\text{\sffamily{LinearSigmoid}}}(\hat{\mathcal{S}}^{2}_{\text{de}}). \\ 
\end{aligned}
\right.
\end{equation}
Here, $\small{\text{\sffamily{3DConvTransGELU}}}^{(E)}(\cdot)$ stands for $E$ 3D inverse convolution blocks, each block consists of a 3D inverse convolution layer and a GELU function, which is used to map the features $\mathcal{X}^{\text{out}}_{\text{en}} \in \mathbb{R}^{\frac{T}{T_p} \times \frac{H}{4H_p} \times \frac{W}{4W_p} \times 4C}$ extracted by encoder to three-channel features $\hat{\mathcal{S}}^{1}_{\text{de}} \in \mathbb{R}^{T \times \frac{H}{4H_p} \times \frac{W}{4W_p} \times 3}$. $\small{\text{\sffamily{Reshape}}}(\cdot)$ stands for the dimensional reshaping operation, which reshapes 4-order $T \times \frac{H}{4H_p} \times \frac{W}{4W_p} \times 3$ into 2-order $T \times (\frac{H}{4H_p} \times \frac{W}{4W_p} \times 3)$. $\small{\text{\sffamily{LinearGELU}}}^{(L)}(\cdot)$ stands for $L$ linear blocks, each block consists of a linear layer and a GELU function, which is used to linearly map reshaped $\hat{\mathcal{S}}^{1}_{\text{de}}$ to low-dimensional features $\hat{\mathcal{S}}^{2}_{\text{de}}$. $\small{\text{\sffamily{LinearSigmoid}}}(\cdot)$ stands for a linear projection block, which consists of a linear layer and a sigmiod activation function. $\small{\text{\sffamily{LinearSigmoid}}}(\cdot)$ linearly projects $\hat{\mathcal{S}}^{2}_{\text{de}}$ as the future SOR $\hat{{\rm P}}^{*}_{T+1:T+K}$, which is used to assist SUs in making advance decisions for DSA. The 3D-SwinLinear training process is illustrated in Algorithm \ref{alg:3D-SwinLinear}. The $\bm{{\rm b}}$ is represented by the same letter as in 3D-SwinSTB. The loss function and optimizer are implemented by (\ref{loss}) and (\ref{eq25}), respectively. The reason why we design a dedicated network prediction SOR instead of calculating SOR with the spectrogram predicted by 3D-SwinSTB is in Appendix B.
\begin{algorithm}[t]
	\caption{TL-Based Training Algorithm for Different Spectrum Services} 
	\label{alg:4}
	\hspace*{0.04in}{\textbf{Initialization}: Load the pre-trained model \\ $\{{D}^{\text{3D}}_{{\bm{\beta}^{*}}^{\text{3D}}}(S_{{\bm{\alpha}^{*}}^{\text{3D}}}(\cdot)), {D}^{\text{SOR}}_{{\bm{\beta}^{*}}^{\text{SOR}}}(S_{{\bm{\alpha}^{*}}^{\text{SOR}}}(\cdot))\}$.}
	
	\begin{algorithmic}[1]
		\STATE \textbf{Input}: The target spectrum service dataset $\mathcal{X}_T$;
		\STATE \textbf{while} $\mathcal{L}$ decreases by less than $\vartheta_{\text{per}}$ \% in $n_{\text{ep}}$ epochs \textbf{do}
		\STATE \quad $\mathcal{Y}^{\text{3D}}_T = {D}^{\text{3D}}_{{\bm{\beta}^{*}}^{\text{3D}}}(S_{{\bm{\alpha}^{*}}^{\text{3D}}}(\mathcal{X}_T))$;\\
		\STATE \quad Compute loss function $\mathcal{L}$ by (\ref{loss}); \\
		\STATE \quad Fine-tune $\{{\bm{\alpha}^{*}}^{\text{3D}}, {\bm{\beta}^{*}}^{\text{3D}}\}$ by optimizer with (\ref{eq25});\\
		\STATE \textbf{end while}
		\STATE \textbf{while} $\mathcal{L}$ decreases by less than $\vartheta_{\text{per}}$ \% in $n_{\text{ep}}$ epochs \textbf{do}
		\STATE \quad $\mathcal{Y}^{\text{SOR}}_T = {D}^{\text{SOR}}_{{\bm{\beta}^{*}}^{\text{SOR}}}(S_{{\bm{\alpha}^{*}}^{\text{SOR}}}(\mathcal{X}_T))$;\\
		\STATE \quad Compute loss function $\mathcal{L}$ by (\ref{loss}); \\
		\STATE \quad Fine-tune $\{{\bm{\alpha}^{*}}^{\text{SOR}}, {\bm{\beta}^{*}}^{\text{SOR}}\}$ by optimizer with (\ref{eq25});\\
		\STATE \textbf{end while}
		\STATE  Obtain updated parameters$\{{{\bm{\alpha}}_{T}^{*}}^{\text{3D}}, {{\bm{\beta}}_{T}^{*}}^{\text{3D}}, {{\bm{\alpha}}_{T}^{*}}^{\text{SOR}}, {{\bm{\beta}}_{T}^{*}}^{\text{SOR}}\}$;\\
		\STATE \textbf{Output}: The re-trained network \\
		\STATE $\{{D}^{\text{3D}}_{{\bm{\beta}_{T}^{*}}^{\text{3D}}}(S_{{\bm{\alpha}_{T}^{*}}^{\text{3D}}}(\cdot)), {D}^{\text{SOR}}_{{\bm{\beta}_{T}^{*}}^{\text{SOR}}}(S_{{\bm{\alpha}_{T}^{*}}^{\text{SOR}}}(\cdot))\}$.	
	\end{algorithmic}
\end{algorithm}

\subsection{TL for Different Spectrum Sevices}\label{5.3}
In the real-world wireless systems, different spectrum services result in varying training data. However, retraining the proposed method to meet the requirements of different spectrum services introduces additional costs (such as training time, memory, and GPU resources). TL involves repurposing a model that has been trained for a specific task and adapting it to perform effectively on a different but related task \cite{9723467}. This methodology is grounded in the idea that the knowledge acquired while solving one problem can be utilized to enhance the learning and performance of the model on a distinct task. Unlike the conventional approach of training a model from scratch for the new task, TL leverages the information accumulated during the training of a pre-existing, well-trained model. Therefore, we apply the TL for solving the cross-spectrum service (also be considered cross-band) prediction problem. According to the idea of the TL, we first introduce the domain, denoted as $\mathcal{D} = \{\mathcal{F}, P(\mathcal{X})\}$ and the task $\mathcal{T} = \{\mathcal{Y}, f(\cdot)\}$. Here, $\mathcal{F}$ is the feature set, $P(\mathcal{X})$ is the data distribution, $\mathcal{Y}$ is the label set, and $f(\cdot)$ is the prediction function. In cross-spectrum service prediction, given a source service domain $\mathcal{D}_S$, a source service prediction task $\mathcal{T}_S$, a target service domain $\mathcal{D}_T$, and a target service prediction task $\mathcal{T}_T$, TL is used to improve the learning of $\mathcal{T}_T$ using the knowledge in $\mathcal{D}_T$, $\mathcal{D}_S$, and $\mathcal{T}_S$, $\mathcal{D}_S \neq \mathcal{D}_T$, and $\mathcal{T}_S \neq \mathcal{T}_T$. The cross-spectrum sevice prediction TL algorithm can be formulated as \cite{9723467}
\begin{equation}\label{Eq_TL}
\begin{aligned}
{\rm arg}& \mathop {\rm min} \limits_{f_T(n)} \mathcal{L}(f_T(\mathcal{X}_T(n), \mathcal{D}_S(n), \mathcal{T}_S(n)), \mathcal{Y}_T(n))\\
&\text{s.t.} \quad \mathcal{D}_S \neq \mathcal{D}_T \quad \text{and} \quad \mathcal{T}_S \neq \mathcal{T}_T,
\end{aligned}
\end{equation}
where $f_T(n)$ is the target spectrum service prediction function, $f_T(\mathcal{X}_T(n), \mathcal{D}_S(n), \mathcal{T}_S(n))$ is the prediction result of $\mathcal{X}_T(n)$ ($n= 1, \cdots, N_T$, $N_T$ is the size of target instance) with the assistance of $\mathcal{D}_S(n)$ and $\mathcal{T}_S(n)$, and $\mathcal{X}_T(n)$ and $\mathcal{Y}_T(n)$ are the input and output labels of target instance, respectively. Note that the higher similarity between $\mathcal{D}_S$ and $\mathcal{D}_T$ results in better TL performance. The maximum mean discrepancy (MMD) \cite{6847217} is used to describe this similarity, which measures the distance between two domain distributions in a regenerated Hilbert space. The MMD of $\mathcal{D}_S$ and $\mathcal{D}_T$ is calculated by
\begin{equation}\label{MMD}
\text{MMD}(\mathcal{D}_S, \mathcal{D}_T) = \parallel \dfrac{1}{n_s}\sum_{i=1}^{n_s} \varphi(x_i^s) - \dfrac{1}{n_t}\sum_{i=1}^{n_t} \varphi(x_i^t) \parallel^{2},
\end{equation}
where $\varphi(\cdot)$ stands for the feature space mapping function.

The training process pseudocode of adopting TL for our two methods is given in Algorithm \ref{alg:4}. We first loads the pre-trained model $D_{{\bm{\beta}}^{*}}(S_{{\bm{\alpha}}^{*}}(\cdot))$ (i.e.,${D}^{\text{3D}}_{{\bm{\beta}^{*}}^{\text{3D}}}(S_{{\bm{\alpha}^{*}}^{\text{3D}}}(\cdot))$, ${D}^{\text{SOR}}_{{\bm{\beta}^{*}}^{\text{SOR}}}(S_{{\bm{\alpha}^{*}}^{\text{SOR}}}(\cdot))$) based on historical data $\bm{{\rm X}}_{1:T}$ for a source spectrum service. For applications with different spectrum services, we only need to fine-tune the network parameters by (\ref{Eq_TL}) to suit the target spectrum service. If the spectrum service is entirely different, that is, with a larger MMD in (\ref{MMD}), the pre-trained model can still reduce time consumption since the weights of some layers in the pre-trained model can be reused in the new model, even if most layers need redesigning.
\begin{figure}
	\centering
	\subfigure[Sensor position]{
		\label{map}\includegraphics[width=30mm,height=36mm]{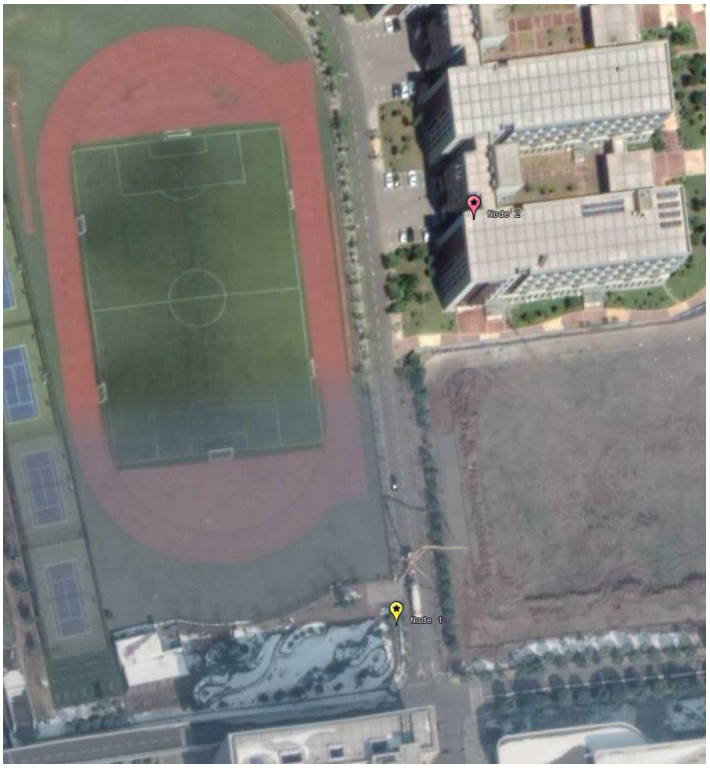}}
	\subfigure[Actual sensor]{
		\label{act}\includegraphics[width=30mm,height=36mm]{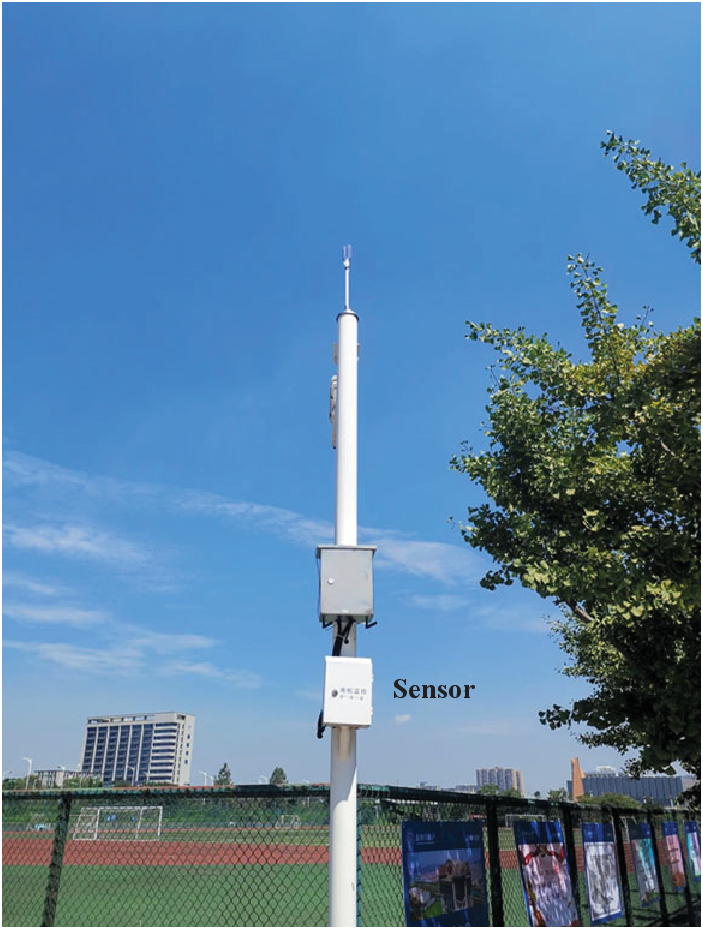}}
	\caption{Explanation of spectrum measurement nodes.}
	\label{sensor}	
\end{figure}

\section{Numerical Results}\label{sec6}
\subsection{Experimental Setup}\label{setup}
\textit{1) Datasets:} This paper adopts three real-world spectrum datasets, including the frequency-modulated (FM) dataset, the long-term evolution (LTE) dataset, and the cross-validation dataset, which are used to analyze the predictive performance, TL performance, and cross-validation of the proposed methods, respectively. The data type and sampling interval for all datasets are I/Q signals and 1 second, respectively. Firstly, the bandwidth of FM dataset and LTE dataset are 90 MHz-110 MHz and 690 MHz-710 MHz, respectively. They are obtained through a spectrum measurement node (located at [118.7905 (east longitude), 31.9378 (northern latitude), 12.10 (altitude)], see Fig. \ref{map} yellow point and Fig. \ref{act} for the actual sensor) deployed at the Jiangjun Road campus of Nanjing University of Aeronautics and Astronautics (NUAA), Nanjing, China. The FM dataset is collected from 17:20 on Sep. 23rd, 2022, to 20:20 on Sep. 23rd, 2022. We preprocess the FM I/Q data collected every second to a spectrogram via the STFT. The STFT configurations are: the sampling frequency is 125 MHz, the descending sampling coefficient is 4, the STFT number is 32508, the center frequency is 99 MHz, and the length-window is 256. We split the FM dataset into the training set (7200 samples with 17:20-19:20), validation set (1800 samples with 19:20-19:50), and test set (1800 samples with 19:50-20:20) with a 4:1:1 ratio in chronological order. The principle of chronological splitting ensures that the model does not contain previously learned data during the testing phase. Secondly, the LTE dataset is collected from 17:52 on May 2nd, 2023, to 18:32 on May 2nd, 2023. The STFT number and the center frequency are 16254 and 700 MHz, respectively. Other settings remain the same as the FM dataset. Thirdly, the cross-validation dataset is obtained from another spectrum measurement node (located at [118.7907 (east longitude), 31.9386 (northern latitude), 36.80 (altitude)], see Fig. \ref{map} pink point) deployed at NUAA with a time range from 22:14 on Jul. 10, 2024, to 00:04 on Jul. 11, 2024. Other settings remain the same as the FM dataset. All datasets are obtained at this repository: https://github.com/pgl1234/Real-world-Spectrum.

\textit{2) Implementation details:} We run all the experiments on a PC with 3.30 GHz Intel Core i9-10940X CPU, NVIDIA GTX 3090Ti graphic, and 64 GB RAM using the Pytorch 1.8.0. The size of spectrogram is $H \times W = 256 \times 256$. The architecture hyperparameters of the 3D-SwinSTB are (1) encoder: $C=96$, 3D Swin Transformer block numbers = $\{2, 4, 2\}$, head numbers = $\{4, 8, 16\}$, (2) the layer numbers of bottleneck layer are 2, and (3) predictor: $C=96$, 3D Swin Transformer block numbers = $\{2, 4, 2\}$, head numbers = $\{16, 8, 4\}$. The patch size and window size of 3D Swin Transformer are $(T_p, H_p, W_p) = (2,4,4)$ and $(P, M, M) = (2,7,7)$, respectively. The loss function is the MSE. The stopping criterion parameters $\vartheta_{\text{per}}=0.01$, $n_{\text{ep}}=4$. The number of train epochs is set as 20. The batch size is 1. The early stop training method is used for our model, and the value of patience is 4. The AdamW optimizer is used to improve the convergence rate of the model. The learning rate is 0.001.

\textit{3) Baselines:} We make a comparison to several state-of-the-art methods, including the dual CNN and GRU (DCG) \cite{yu2020spectrum}, spatial-temporal-spectral prediction network (using convLSTM) \cite{9296309}, spatio-temporal spectrum load prediction model (named NN-ResNet) \cite{9664805}, and stacked autoencoder (SAE)-based spectrum prediction model (named SAE-TSS) \cite{10064355}.

\begin{itemize} 
	\item \textbf{DCG} \cite{yu2020spectrum}: A converged network, which uses a dual CNN as a spectral feature extractor and uses a GRU to mine the long-term temporal features.
	\item \textbf{ConvLSTM} \cite{9296309}: A converged network, which uses three ConvLSTM to model the temporal, spectral, and spatial dependencies of the spectrum data. 
	\item \textbf{NN-ResNet} \cite{9664805}: A converged network, which combines both CNN and ResNet to predict spatio-temporal spectrum usage
	of the region. The ResNet's skip connection helps to alleviate the gradient vanishing problem.
	\item \textbf{SAE-TSS} \cite{10064355}: A converged network, which uses SAE layer by layer to extract the feature information of temporal-spectral-spatial spectrum data while reducing the data dimension. A predictor combining both CNN and stacked Bi-LSTM is then used to capture these temporal-spectral-spatial features. 
\end{itemize}
\begin{figure*}[t]
	\centering
	\subfigure[Frame-wise MSE]{
		\label{Fig:MSE}\includegraphics[width=70mm,height=60mm]{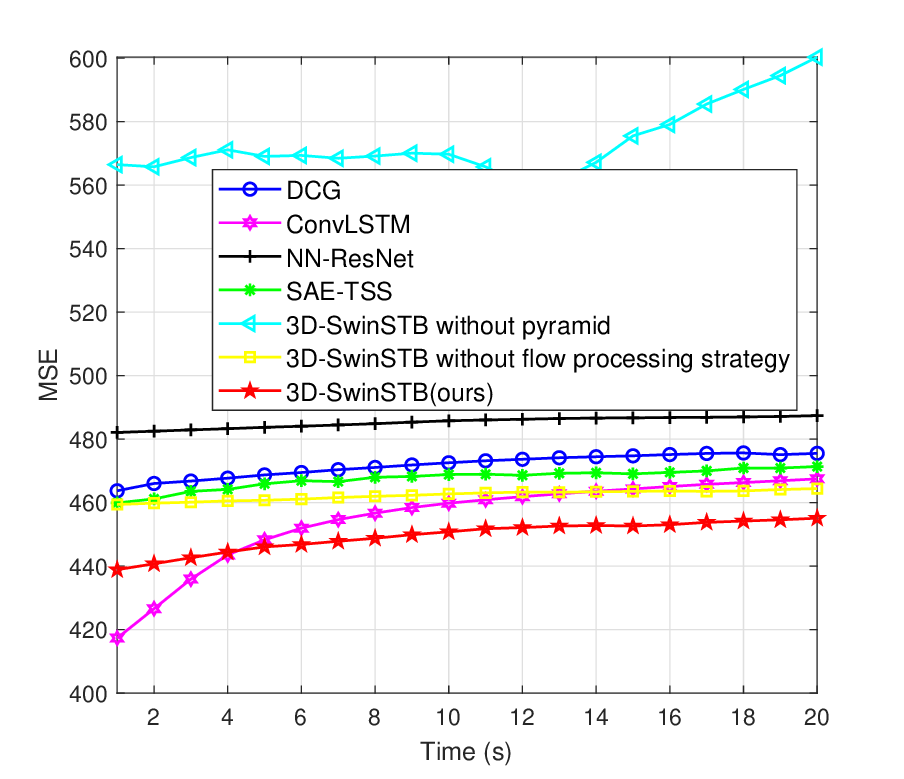}}
	\subfigure[Frame-wise SSIM]{
		\label{Fig:SSIM}\includegraphics[width=70mm,height=60mm]{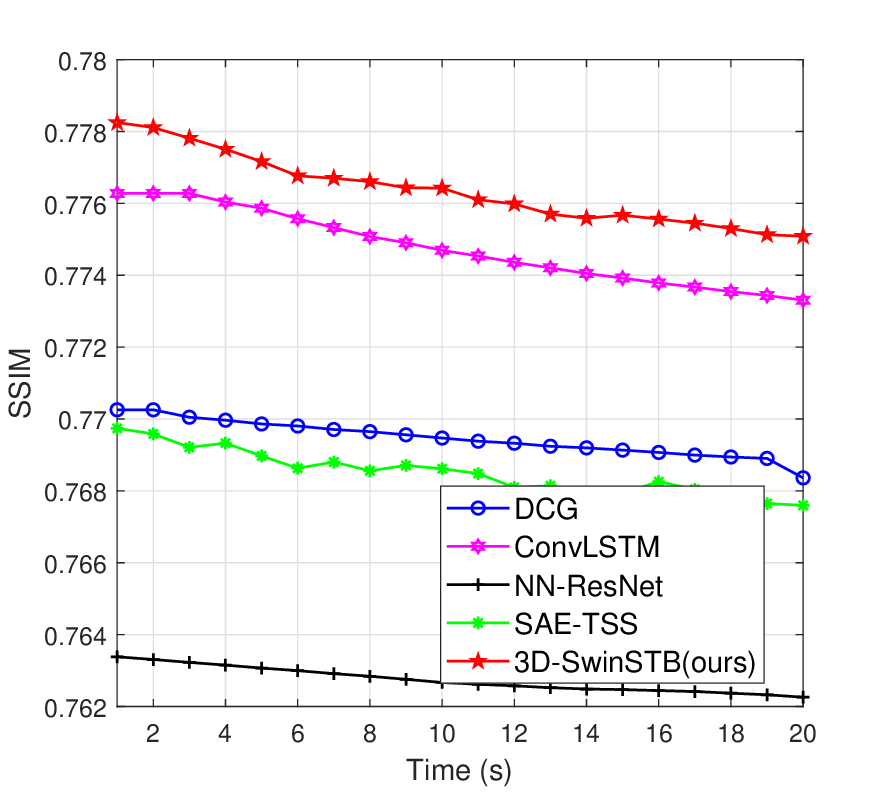}}
	\subfigure[Frame-wise PSNR]{
		\label{Fig:PSNR}\includegraphics[width=70mm,height=60mm]{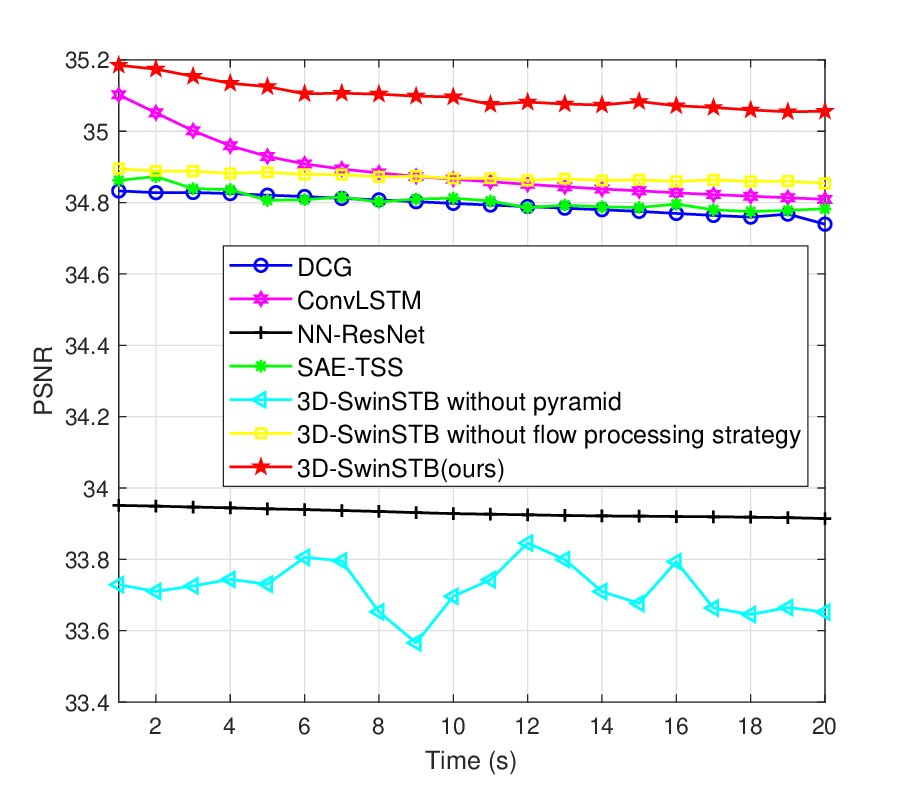}}
	\subfigure[Frame-wise LPIPS]{
		\label{Fig:Lpips}\includegraphics[width=70mm,height=60mm]{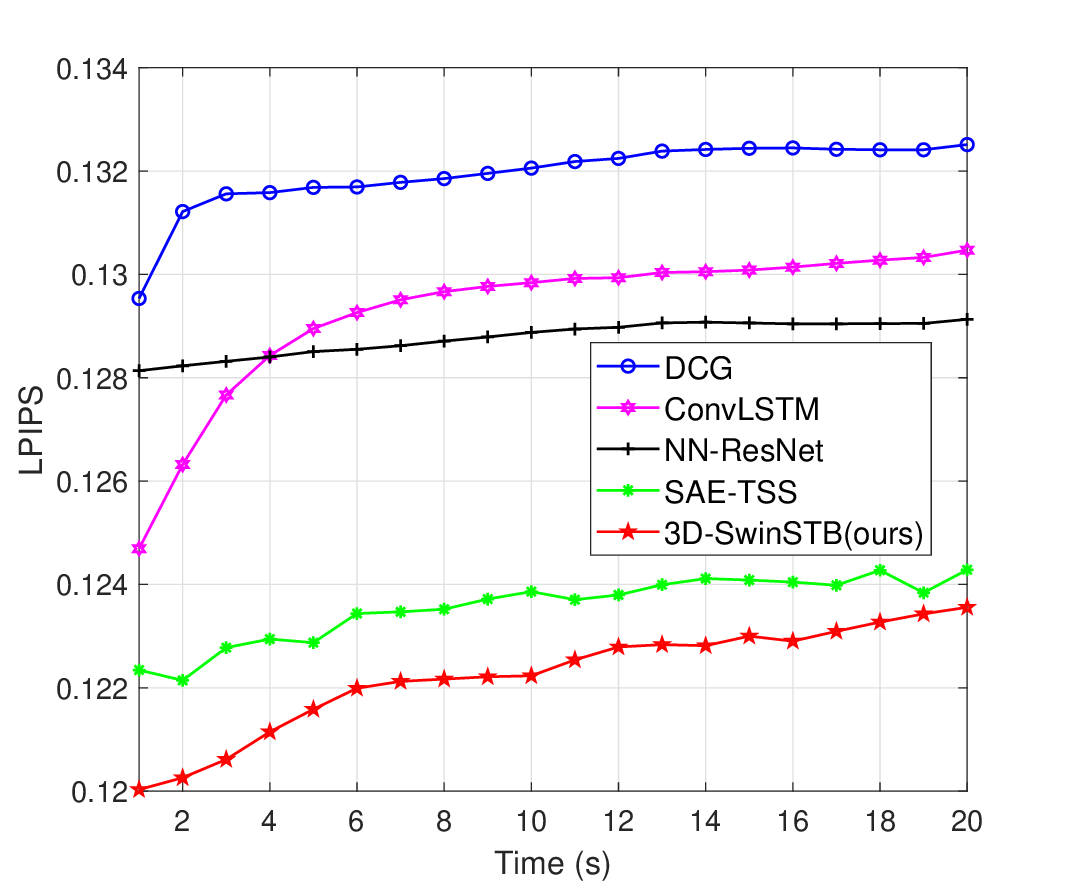}}
	\caption{Results of frame-wise MSE, SSIM, PSNR, and LPIPS comparison of proposed 3D-SwinSTB with all baselines under the prediction of 20 frames.}
	\label{Fig7}	
\end{figure*}

\textit{4) Evaluation metrics:} For 3D spectrum prediction, the performance depends on the predicted spectrogram quality. Thus, we use the MSE, peak signal-to-noise ratio (PSNR), structural similarity (SSIM) \cite{1284395}, and learned perceptual image patch similarity (LPIPS) \cite{zhang2018unreasonable} as evaluation metrics, which are
\begin{equation}
\label{mse}
\text{MSE}(\hat{X}^{*}_{k}, X^{*}_{k})=\sum_{i=0}^{h-1}\sum_{j=0}^{w-1}[\hat{X}^{*}_{k}(i,j)-X^{*}_{k}(i,j)]^2,
\end{equation}
\begin{equation}
\label{PSNR}
\text{PSNR}(\hat{X}^{*}_{k}, X^{*}_{k})= 10 \cdot \text{log}_{10}(\dfrac{\text{MAX}^2_I}{\text{MSE}(\hat{X}^{*}_{k}, X^{*}_{k})}),
\end{equation}
\begin{equation}
\label{SSIM}
\text{SSIM}(\hat{X}^{*}_{k}, X^{*}_{k})= \dfrac{(2\mu_{\hat{X}^{*}_{k}}\mu_{X^{*}_{k}} + c_1)(2\sigma_{\hat{X}^{*}_{k} X^{*}_{k}}+c_2)}{(\mu^2_{\hat{X}^{*}_{k}}+\mu^2_{X^{*}_{k}}+c_1)(\sigma^2_{\hat{X}^{*}_{k}} + \sigma^2_{\hat{X}^{*}_{k}}+c_2)},
\end{equation}
\begin{equation}
\label{LPIPS}
\text{LPIPS}(\hat{X}^{*}_{k}, X^{*}_{k})= \dfrac{1}{N}\sum_{i=1}^{N}\|\phi_i(\hat{X}^{*}_{k})-\phi_i(X^{*}_{k})\|_2.
\end{equation}
Here, $\hat{X}^{*}_{k}$ and $X^{*}_{k}$ are predicted data and true data with $k$th frame, respectively, and $h$ and $w$ are the height and width of the image, respectively, $\text{MAX}_I = 2^8-1=255$. Further, $\mu_{\hat{X}^{*}_{k}}$ and $\mu_{X^{*}_{k}}$ are the averages of $\hat{X}^{*}_{k}$ and $X^{*}_{k}$, respectively, $\sigma_{\hat{X}^{*}_{k}}$ and $\sigma_{\hat{X}^{*}_{k}}$ are the variances of $\hat{X}^{*}_{k}$ and $X^{*}_{k}$, respectively, and $c_1 = (\rho_1L)^2$ and $c_2 = (\rho_2L)^2$ are the constants used to maintain stability, where $L$ is the range of pixel values, $\rho_1 = 0.01$ and $\rho_2 = 0.03$. Further, $\phi_i(\cdot)$ indicates the output of the $i$ feature block in the LPIPS network and $N$ is the number of feature blocks. We use a threshold decision to evaluate SOR predictive performance. Specifically, we conduct $N_{\text{time}}$ experiments for $K$ consecutive frames SOR prediction on the testset, where each frame corresponds to a SOR value and $N_{\text{time}}$ is set to 100. We then set a threshold of $\lambda$ for the error $e_{\text{err}}$ between the predicted SOR value and the true value for each frame. If $e_{\text{\text{err}}} \leq \lambda$, the prediction is considered correct. Otherwise, it is incorrect. So, the prediction accuracy is $T_{e_{\text{err}} \leq \lambda}/(K \times N_{\text{time}})$, where $T_{e_{\text{err}} \leq \lambda}$ is the number of correct predictions.

\subsection{Comparison to State-of-the-Art}\label{Comparison}
In this subsection, we make a comparison to several state-of-the-art methods (baselines in Section \ref{setup}). To achieve a fair comparison, the same hyperparameters are used, except for the network structure. The trained hyper-parameters of the baseline methods remain in the same configuration as the proposed 3D-SwinSTB. Fig. \ref{Fig7} provides the MSE, SSIM, PSNR, and LPIPS with frame-wise comparison results for predicting 20 frames. From Fig. \ref{Fig7}, the proposed 3D-SwinSTB significantly outperforms the baseline methods across all metrics. For example, in Fig. \ref{Fig:MSE}, we can see the 3D-SwinSTB has a MSE decrease of 4.51\% (from 475.6372 to 454.1826) to DCG at 18th frame. In Fig. \ref{Fig:SSIM}, 3D-SwinSTB has a SSIM improvement of 0.23\% (from 0.7735 to 0.7753) to ConvLSTM at 18th frame. In Fig. \ref{Fig:PSNR}, 3D-SwinSTB has a PSNR improvement of 3.45\% (from 33.9343 to 35.1039) to NN-ResNet with 8th frame. In Fig. \ref{Fig:Lpips}, 3D-SwinSTB has a LPIPS decrease of 1.05\% (from 0.1235 to 0.1222) to SAE-TSS with 8th frame. These results indicate that the designed flow processing strategy and pyramid structure capture the spatiotemporal behavior dependencies of the PUs across different frequency bands far better than the baseline. Note that the MSE of ConvLSTM is lower than that of our method when the number of predicted frames is below 4. This indicates that ConvLSTM is proficient at capturing short-term changes in user behavior. The spectrum users from different frequency bands exhibit complex and intertwined usage patterns over time. Our method's flow processing strategy enables the learning of long-term spectrum usage patterns.

Fig. \ref{Fig:MSE} shows the MSE of the 3D-SwinSTB is significantly lower than that of the 3D-SwinSTB without the pyramid and flow processing strategy. Fig. \ref{Fig:PSNR} shows the PSNR of the 3D-SwinSTB is significantly higher than that of them. These results indicate performance gains in learning spectrum usage patterns for each design in our method, including the pyramid and flow processing strategy. Note that the performance gain brought by the proposed pyramid surpasses that of the proposed flow processing strategy, and the 3D-SwinSTB without the pyramid exhibits a poorer prediction stability.
\begin{figure*}
	\centering
	\subfigure[Frame-wise MSE]{
		\label{Fig:MSE_other}\includegraphics[width=70mm,height=62mm]{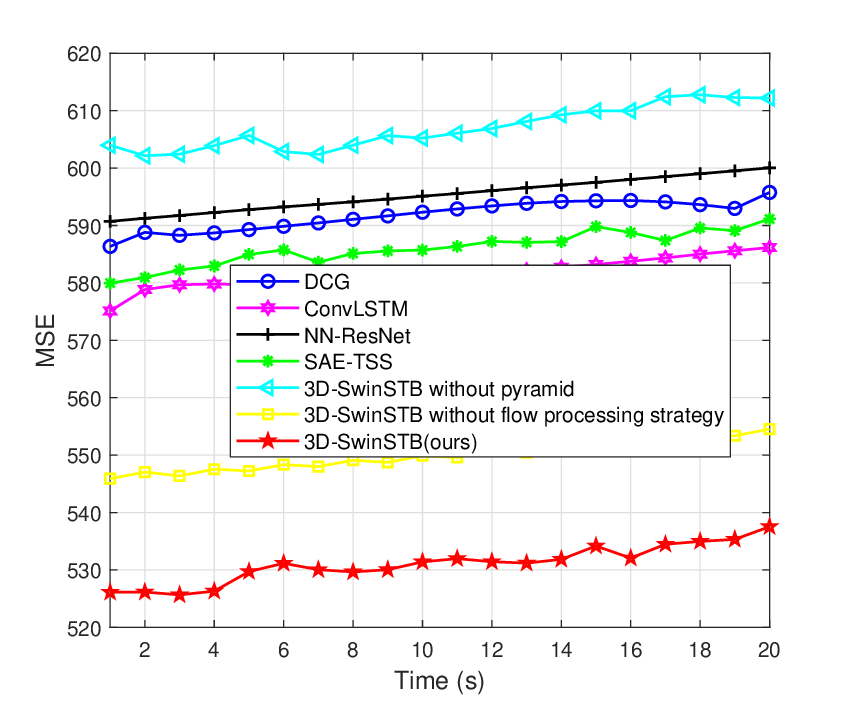}}
	\subfigure[Frame-wise PSNR]{
		\label{Fig:PSNR_other}\includegraphics[width=70mm,height=62mm]{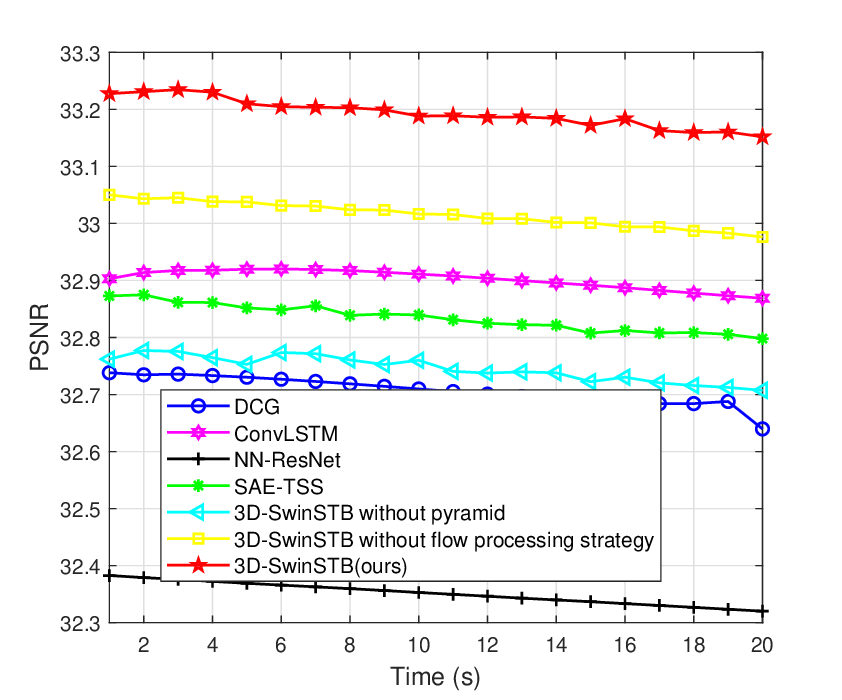}}
	\caption{Results of frame-wise MSE and PSNR comparison of proposed 3D-SwinSTB with all baselines using a cross-validation dataset.}
	\label{Fig7_other}	
\end{figure*}
\begin{figure}
	\centerline{\includegraphics[width=70mm,height=60mm]{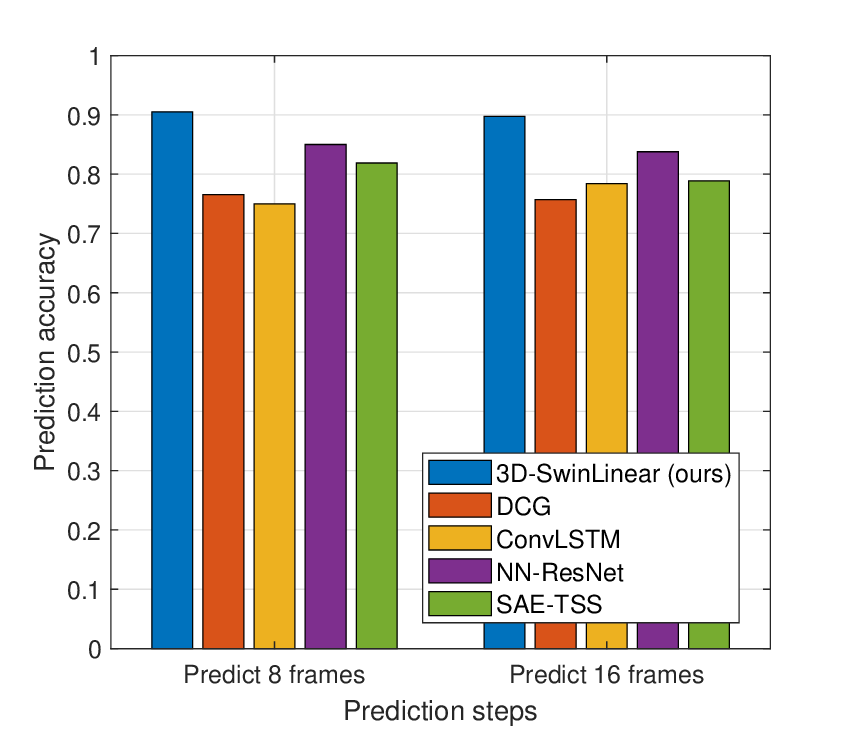}}
	\caption{Comparison of prediction accuracy. Since there is currently no image-based effort to predict the SOR, all baselines are redesigned to embed the proposed predictor for comparison.}
	\label{Fig:Acc}
\end{figure}

To verify the generalization capability and applicability of the proposed method across different times and locations, Fig. \ref{Fig7_other} shows frame-wise MSE and PSNR results of the proposed 3D-SwinSTB compared with the baseline methods on a cross-validation dataset. Firstly, the prediction performance of 3D-SwinSTB is significantly better than that of all baseline methods. For example, in Fig. \ref{Fig:MSE_other}, the 3D-SwinSTB has a MSE decrease of 10.44\% (593.4100 $\rightarrow$ 531.4336) to DCG at 12th frame. In Fig. \ref{Fig:PSNR_other}, the 3D-SwinSTB has a PSNR improvement of 2.58\% (32.3268 $\rightarrow$ 33.1593) to NN-ResNet at 18th frame. Secondly, the proposed pyramid and flow processing strategy are also verified to be effective. For example, in Fig. \ref{Fig:MSE_other}, the 3D-SwinSTB has a MSE decrease of 12.43\% (606.8934 $\rightarrow$ 531.4336) to 3D-SwinSTB without pyramid at 12th frame. In Fig. \ref{Fig:PSNR_other}, the 3D-SwinSTB has a PSNR improvement of 1.36\% (32.7158 $\rightarrow$ 33.1593) to 3D-SwinSTB without flow processing strategy at 18th frame.

Fig. \ref{Fig:Acc} shows the comparison results of prediction accuracy between the proposed 3D-SwinLinear and the baselines. We can see that the accuracy of the 3D-SwinLinear in predicting 8 and 16 frames is maintained at about 90\%. In contrast, all baselines are below 90\%. For example, the accuracy of the 3D-SwinLinear is 18.21\% (from 0.7656 to 0.9050) and 20.67\% (from 0.7500 to 0.9050) higher than that of DCG and ConvLSTM with 8 frames, respectively. Further, the accuracy of the 3D-SwinLinear is 7.13\% (from 0.8378 to 0.8975) and 13.78\% (from 0.7888 to 0.8975) higher than that of NN-ResNet and SAE-TSS with 16 frames, respectively. These results show that our method can help SUs to know the spectrum occupancy information with the best accuracy in advance, so as to achieve efficient DSA.
\begin{figure*}[t]
	\centering
	\subfigure[ ]{
		\label{Fig:TL-MSE}\includegraphics[width=36mm,height=48mm]{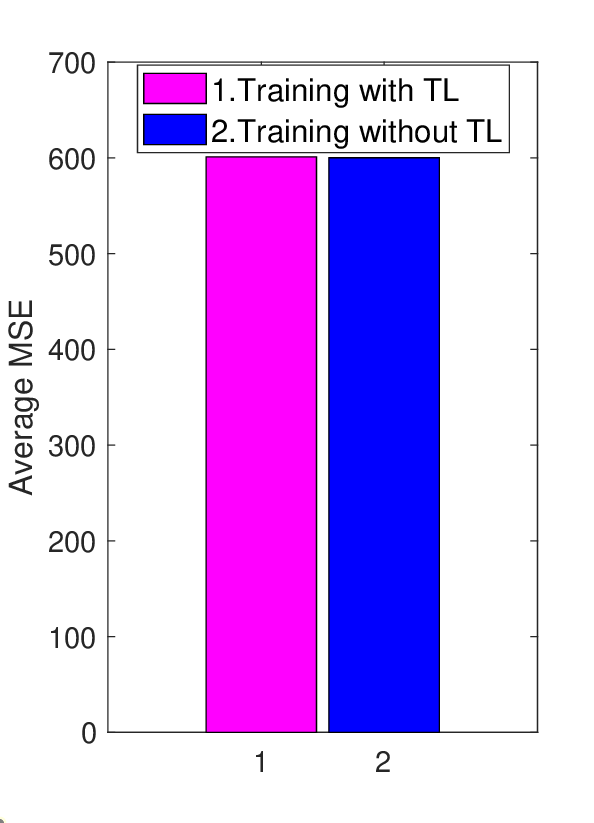}}
	\subfigure[ ]{
		\label{Fig:TL-Time}\includegraphics[width=36mm,height=48mm]{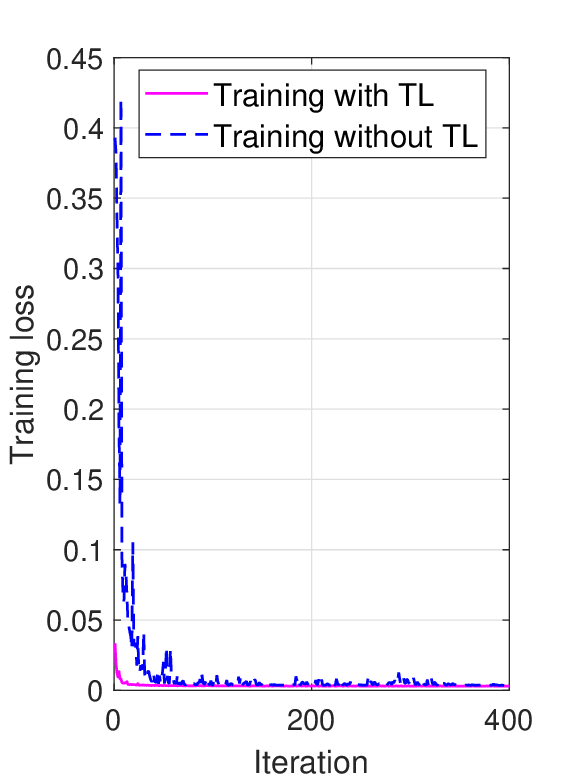}}
	\subfigure[ ]{
		\label{Fig:TL-Loss}\includegraphics[width=36mm,height=48mm]{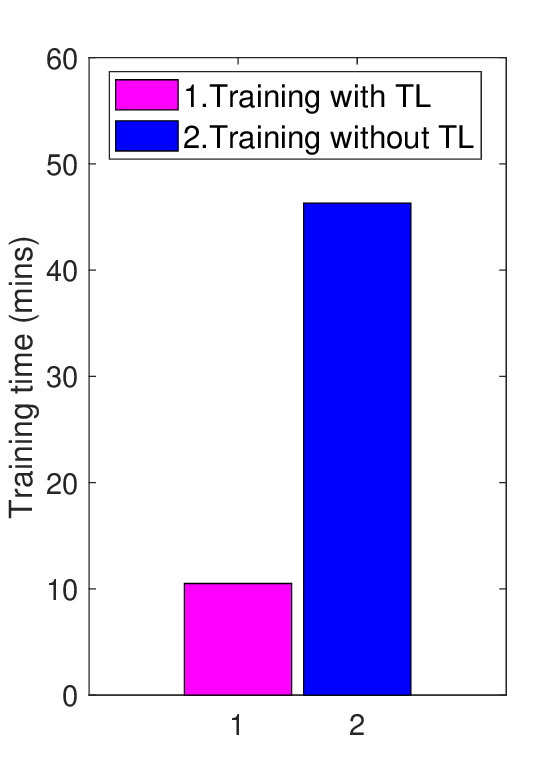}}
	\subfigure[ ]{
		\label{Fig:TL-Learnrate}\includegraphics[width=36mm,height=48mm]{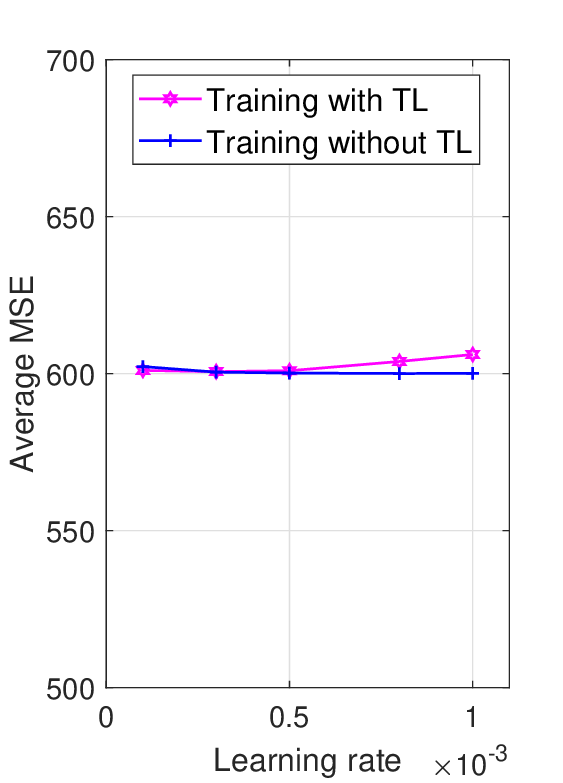}}
	\caption{Performance analysis of TL-aided 3D-SwinSTB with different spectrum services (FM $\rightarrow$ LTE): (a) Comparison of average MSE ; (b) Loss values versus iterations; (c) Comparison of training time; (d) Average MSE versus the learning rate. Here, the average MSE is calculated by predicting 16 frames.}
	\label{Fig:TL}	
\end{figure*}   
\begin{table}
	\renewcommand\arraystretch{1.3}
	\renewcommand\tabcolsep{1.8pt}  
	\centering		
	\caption{Comparison of Model Efficiency}\label{tab:Efficiency}
	\begin{tabular}{ c | c | c | c | c | c | c | c }
		\hline
		\multirow{2}*{Method} & \multirow{2}*{\makecell{Params.\\(MB)}} & \multicolumn{3}{c|}{Input-8-Predict-8} & \multicolumn{3}{c}{Input-16-Predict-16} \\ 
		\cline{3-8}
		& & \makecell{FLOPs \\ (GB)} & \makecell{ATT \\ (mins)} & \makecell{AIT \\ (s)} & \makecell{FLOPs \\ (GB)} & \makecell{ATT \\ (mins)} & \makecell{AIT \\ (s) } \\
		\hline  
		DCG & 3.65 & 15.75 & 72.42 & 2.5000 & 31.5  & 60.57 & 2.0904 \\ 
		ConvLSTM & 9.06 & 109.44 & 63.40 & 2.1863 & 218.88 & 144.89 & 2.3661 \\ 
		NN-ResNet & 14.36 & 70.03 & 65.80 & 2.6934 & 140.06 & 54.33 & 2.5203  \\   
		SAE-TSS & 21.69& 7.97 & 42.64 & 2.0173 & 15.94 & 31.55 & 2.4656 \\  
		\hline 
		\textbf{3D-SwinSTB} & \textbf{16.32} & \textbf{12.88} & \textbf{85.65} & \textbf{2.3966}& \textbf{25.76} & \textbf{104.18} & \textbf{2.1969} \\
		\hline
	\end{tabular}
\end{table}

\subsection{TL for Different Spectrum Services}\label{TL}
In this experiment, we present the performance of TL-based 3D-SwinSTB for diffident spectrum services. The pre-trained model with a FM dataset is migrated to the LTE service to achieve spectrum prediction. Note that the proposed 3D-SwinLinear and 3D-SwinSTB have the same TL process, which is not analyzed in detail.

Fig. \ref{Fig:TL} shows the performance and the training efficiency for TL-based spectrum prediction. The models have the same structure and re-train with the same parameters in each service. From Fig. \ref{Fig:TL-MSE}, the prediction average MSE of TL-based model is comparable to that directly in the new LTE service training model. This demonstrates that the TL-aided 3D-SwinSTB can help the model to accommodate the new requirements of service. From Fig. \ref{Fig:TL-Time} and \ref{Fig:TL-Loss}, compared with the model without TL, the TL-based model can reach convergence quickly in training, and the training time decreases by 77.32\% (from 46.3 to 10.5). This shows TL-based model can realize the trade-off between training efficiency and performance. In Fig. \ref{Fig:TL-Learnrate}, the performance of TL-based 3D-SwinSTB remains similar to that of 3D-SwinSTB without TL across different learning rate settings. This shows that our TL-model has a robustness to hyper-parameters in new spectrum service.

\subsection{Model Efficiency Analysis}\label{Analysis}
In this subsection, we analyze the efficiency of the proposed 3D-SwinSTB and all baselines to provide a comprehensive and fair computational comparison. We consider \textit{input-8-predict-8} and \textit{input-16-predict-16}, which correspond to short-term and long-term predictions. The trained hyper-parameters of all baselines remain in the same configuration as the proposed method. Table \ref{tab:Efficiency} presents the number of parameters, FLOPs, average training time (ATT), and average inference time (AIT) of different methods. Here, we take the average of the two experimental results as the final result. In Table \ref{tab:Efficiency}, the parameters of the proposed 3D-SwinSTB are 12.67 MB (from 3.65 to 16.32), 7.26 MB (from 9.06 to 16.32) and 7.26 MB (from 9.06 to 16.32) higher than those of the previous three comparison methods, respectively, but 5.37 MB (from 21.69 to 16.32) lower than that of SAE-TSS. For the FLOPs, the 3D-SwinSTB is lower than the first three baselines but higher than SAE-TSS. Although the ATT of our method is the highest in the \textit{input-8-predict-8}, the predictive performance of our method is superior to all baselines, as proved in Section \ref{Comparison}. The AIT (inference hundreds of times) of all methods is 2-3s with our method is in the middle level. These results show that our method can perform spectrum monitoring tasks with low complexity and high accuracy to rapidly infer future spectrogram details to help management entities perform downstream monitoring tasks such as anomaly detection.
\begin{table}
	\renewcommand\arraystretch{1}
	\begin{center}		
		\caption{Ablation study for the number of 3D Swin Transformer blocks and channels}\label{tab:3D-Blocks}
		\begin{tabular}{c c c c c}
			\hline  
			\makecell{Number of 3D \\ Transfomer blocks} & $\{2, 2, 2\}$ & $\{2, 4, 2\}$ & $\{2, 6, 2\}$ \\
			\midrule  
			MSE (8th frame) & \textbf{447.4807} & 448.7628 & 450.8983 \\
			MSE (16th frame) & 455.0453 & \textbf{453.0376} & 456.3673 \\
			\midrule
			Number of channels $C$ & 48 & 96 & 128 \\
			\midrule
			MSE (8th frame) & 454.8080 & \textbf{448.7628} & 453.0270 \\  
			MSE (16th frame) & 470.2125 & \textbf{453.0376} & 457.4434 \\
			\hline
		\end{tabular}		
	\end{center}
\end{table}
\begin{table}
	\renewcommand\arraystretch{1}
	\begin{center}		
		\caption{Ablation study for learning rate and patch and window}\label{tab:learning_rate}
		\begin{tabular}{c c c c}
			\hline  
			Learning rate & 0.01 & 0.001 & 0.0001 \\
			\midrule
			MSE (8th frame) & $\times$ & \textbf{448.7628} & 459.8991 \\
			MSE (16th frame) & $\times$ & \textbf{453.0376} & 460.7154 \\  
			\midrule  
			Patch and window & \makecell{$\{2, 2, 2\}$ \\ $\{2, 4, 4\}$} & \makecell{$\{2, 4, 4\}$ \\ $\{2, 7, 7\}$}  & \makecell{$\{4, 4, 4\}$ \\ $\{4, 7, 7\}$} \\
			\midrule
			MSE (8th frame) & \textbf{440.8266} & 448.7628 & 455.1307  \\
			MSE (16th frame) & 455.7650 & \textbf{453.0376} & 467.7145 \\  
			\hline
		\end{tabular}		
	\end{center}
\end{table}

\subsection{Ablation Studies} \label{Ab}
In this subsection, we perform ablation studies to analyze the efficacy of each design choice in the proposed 3D-SwinSTB. We take \textit{input-20-predict-20} as an example. We analyze the performance of each design to keep the other designs the same to ensure a fair comparison.
\begin{figure*}[t]
	\centering
	\subfigure[MSE]{
		\label{Fig:Limit_MSE}\includegraphics[width=35mm,height=45mm]{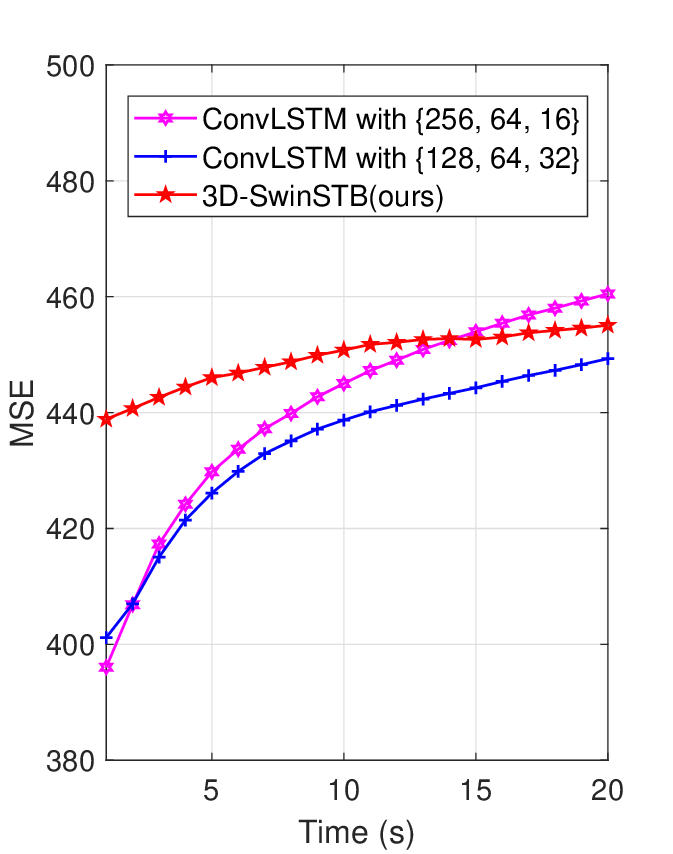}}
	\subfigure[SSIM]{
		\label{Fig:Limit_SSIM}\includegraphics[width=35mm,height=45mm]{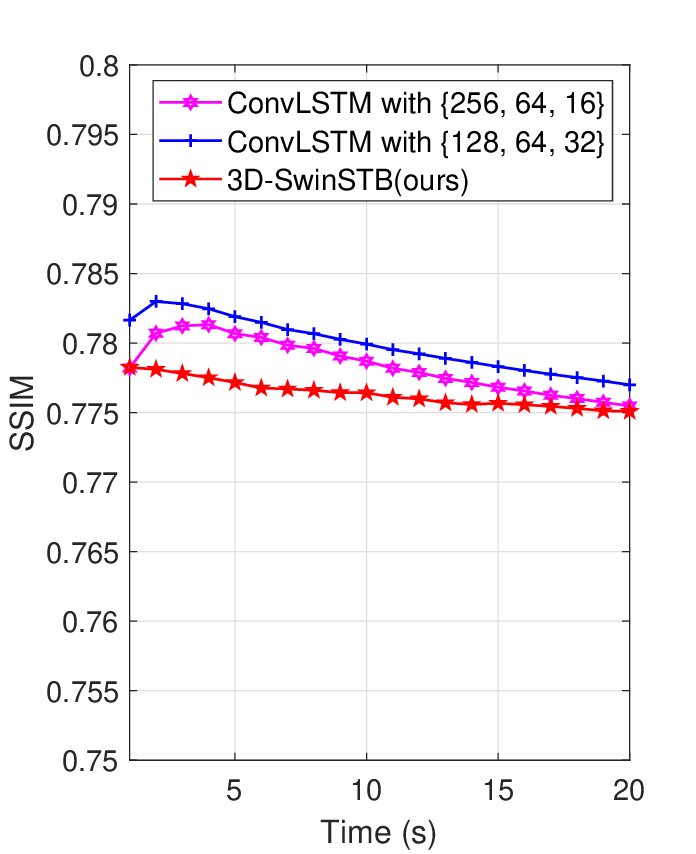}}
	\subfigure[PSNR]{
		\label{Fig:Limit_PSNR}\includegraphics[width=35mm,height=45mm]{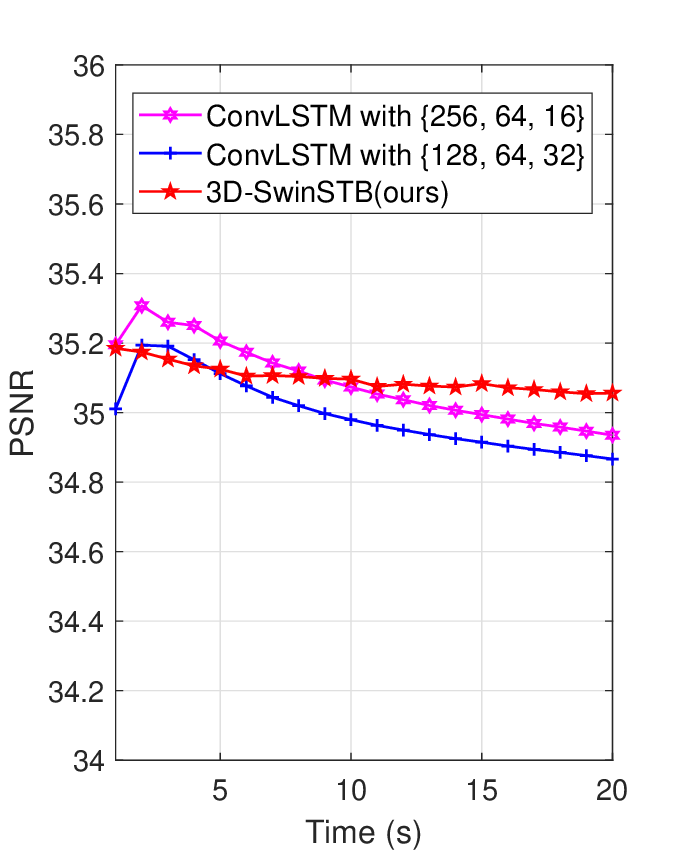}}
	\subfigure[LPIPS]{
		\label{Fig:Limit_LPIPS}\includegraphics[width=35mm,height=45mm]{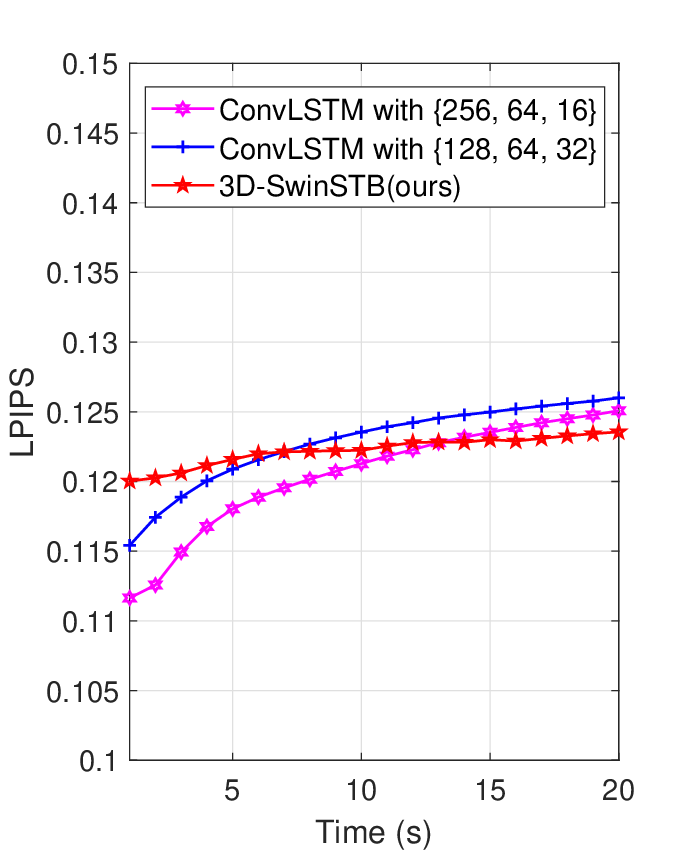}}
	\\
	\subfigure[Param. (M)]{
		\label{Fig:Limit_PARAM}\includegraphics[width=35mm,height=45mm]{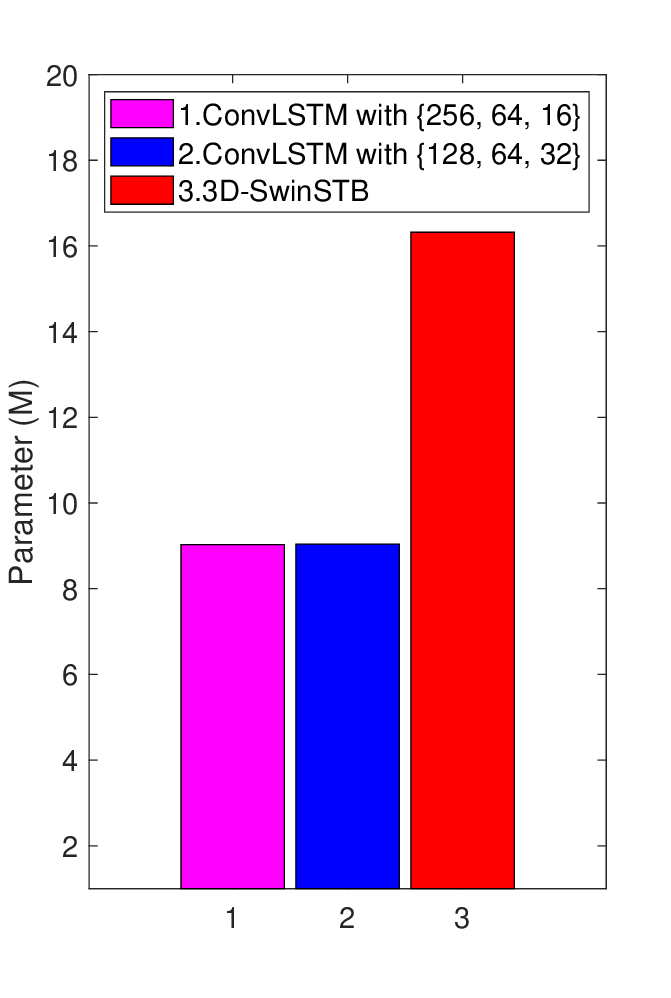}}
	\subfigure[GFlops]{
		\label{Fig:Limit_GFlops}\includegraphics[width=35mm,height=45mm]{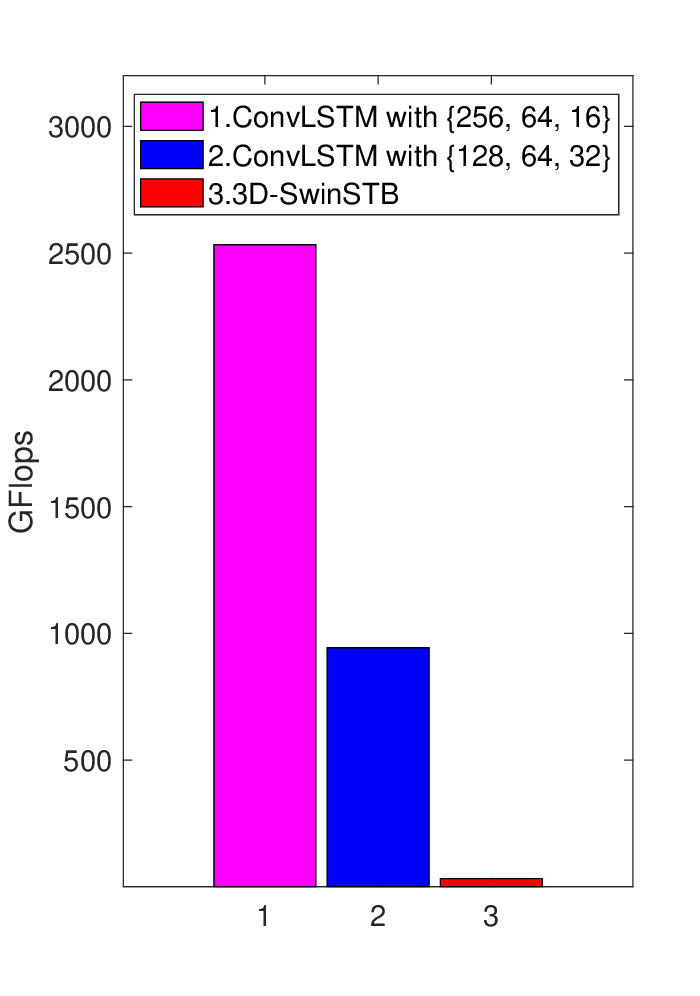}}
	\subfigure[ATT]{
		\label{Fig:Limit_ATT}\includegraphics[width=35mm,height=45mm]{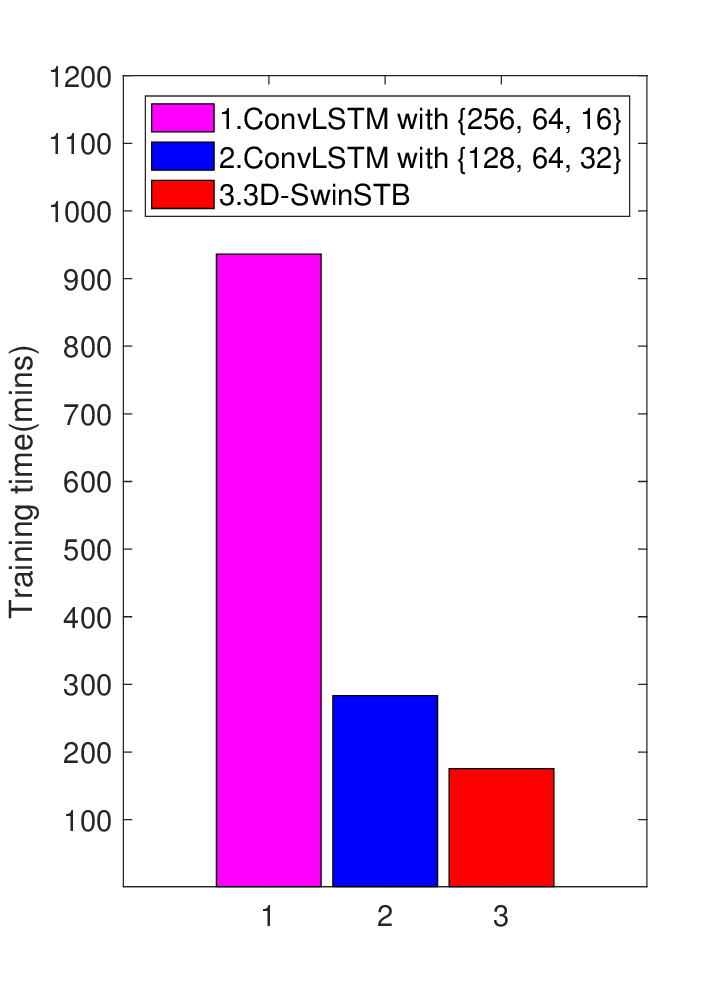}}
	\subfigure[AIT]{
		\label{Fig:Limit_AIT}\includegraphics[width=35mm,height=45mm]{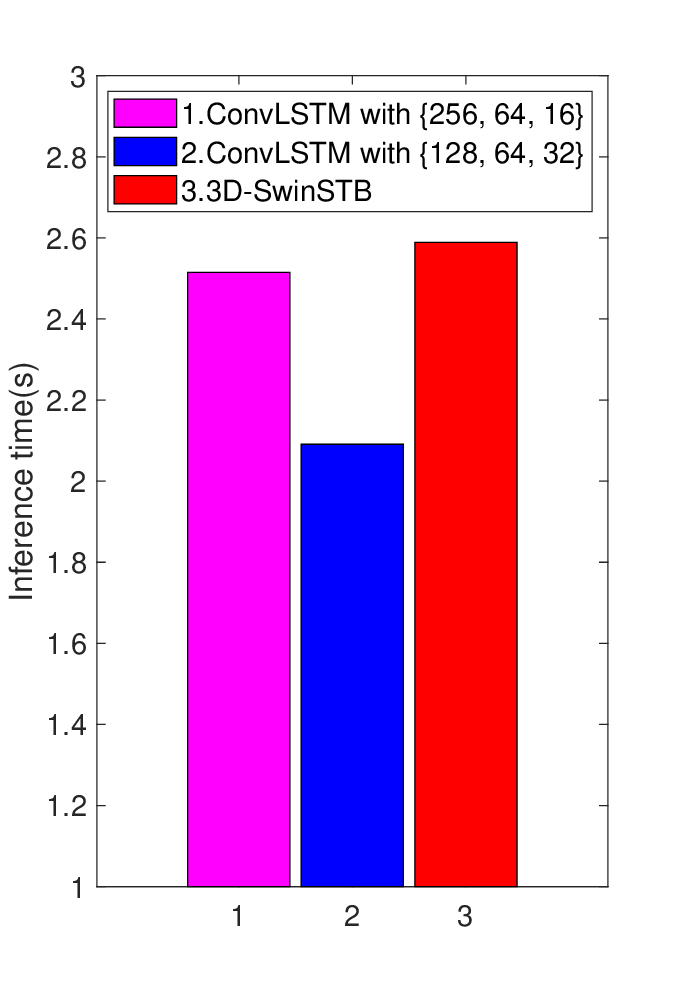}}
	\caption{An example of model limitation analysis. The accuracy performance (includes MSE, SSIM, PSNR, and LPIPS in (a)-(d)) and the model complexity (includes parameter quantity (MB), GFlops, ATT, and AIT in (e)-(h)) of the proposed 3D-SwinSTB are compared with those of ConvLSTM with different network structures under the prediction of 20 frames.}
	\label{Fig:Limit}	
\end{figure*}

\textbf{Number of 3D Swin Transformer blocks.} We conduct experiments for different 3D Swin Transformer block numbers: $\{2, 2, 2\}$, $\{2, 4, 2\}$, and $\{2, 6, 2\}$. From Table \ref{tab:3D-Blocks}, the $\{2, 4, 2\}$ design has a MSE decrease of 2.0077 to the $\{2, 2, 2\}$ and a MSE decrease of 3.3297 to the $\{2, 2, 2\}$ in the 16th frame, respectively. Although $\{2, 4, 2\}$ design has a MSE increase of 2.3 to the $\{2, 2, 2\}$ in 8th frame, it's only by a small margin. Thus, the number of 3D Swin Transformer blocks is set to 2.

\textbf{Number of channels.} The number of channels 48, 96, and 128 are considered for 3D-SwinSTB. From Table \ref{tab:3D-Blocks}, the 96 channels design has an optimal performance. Specifically, the 96 channels design has a MSE decrease of 2.0077 to the 48 channels design in the 8th frame. The 96 channels design has a MSE decrease of 2.0077 to the 128 channels design in the 16th frame. Thus, the channel number is set to 96.

\textbf{Learning rate choices.} We set it to $\{0.01, 0.001, 0.0001\}$, the results are shown in Table \ref{tab:learning_rate} (where $\times$ means out of the error range). The 0.001 design has a MSE decrease of 11.1363 to the 0.0001 design with 8th frame and a MSE decrease of 7.6778 to the 0.0001 design with 16th frame, respectively. When the learning rate is 0.01, it's not counted due to the large MSE. Hence, the learning rate is set to 0.001.

\textbf{Sizes of patch and window.} We set them to $\{\{2, 2, 2\}$, $\{2, 4, 4\}\}$, $\{\{2, 4, 4\}$, $\{2, 7, 7\}\}$, and $\{\{4, 4, 4\}$, $\{4, 7, 7\}\}$, the results are shown in Table \ref{tab:learning_rate}. From Table \ref{tab:learning_rate}, the second design has a MSE increase of 7.9362 to the first design in 8th frame. However, we can see that the second design has a MSE decrease of 2.7274 to the first design and a MSE decrease of 14.6769 to the third design in the 16th frame. In this work, we focus on long-term prediction. Thus, we apply $\{\{2, 4, 4\}$, $\{2, 7, 7\}\}$ as the default sizes of patch and window choices.

\textbf{Other design choices.} The optimal configuration, including design choices such as the number of encoder/decoder layers, optimizer, and head number, is provided in Section \ref{setup}.
\begin{figure}[t]
	\centering
	\subfigure[Frame-wise MSE]{
		\label{Fig:MSE_channel}\includegraphics[width=42mm,height=48mm]{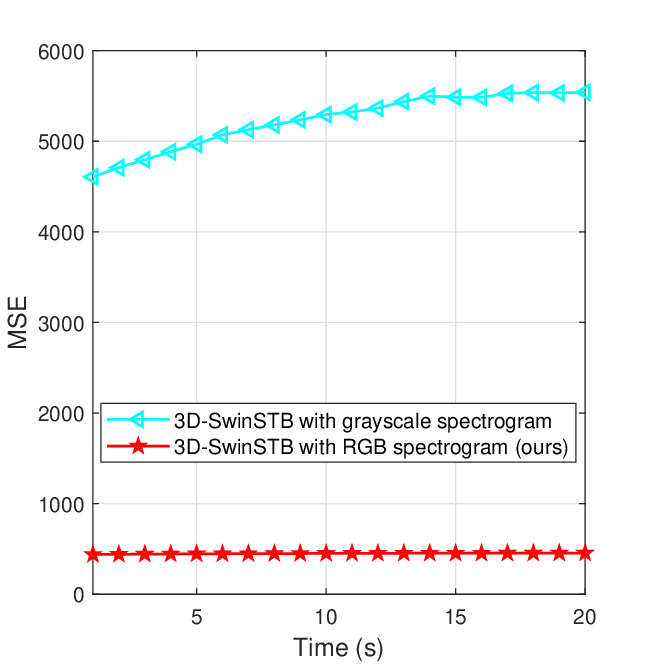}}
	\subfigure[Frame-wise PSNR]{
		\label{Fig:PSNR_channel}\includegraphics[width=42mm,height=48mm]{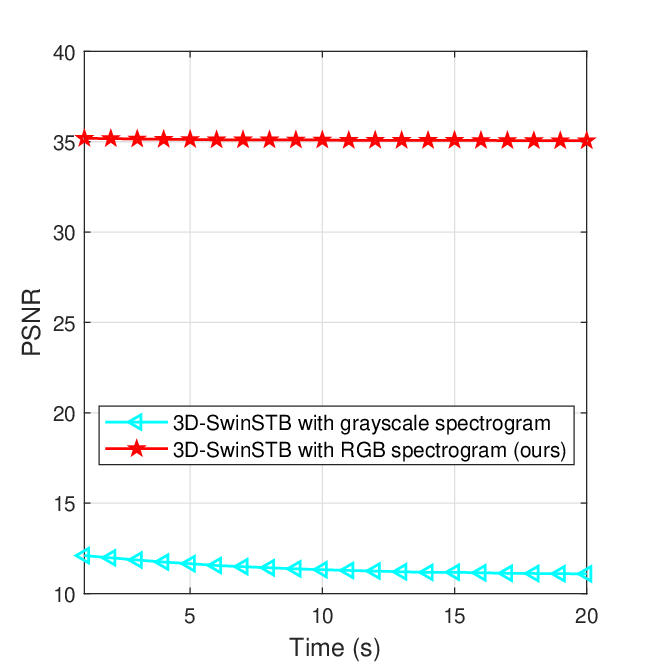}}
	\caption{Results of frame-wise MSE and PSNR comparison of the 3D-SwinSTB using RGB spectrogram and grayscale image, respectively.}
	\label{Figchannel}	
\end{figure}

\subsection{Model Limitations} \label{limit}
In spectrum monitoring and DSA tasks, spectrum management entities need to make fast and accurate decisions, which requires spectrum prediction models to achieve a trade-off between complexity and accuracy. If spectrum management entity has abundant computing resources available, the performance of our method may still have limitations. Fig. \ref{Fig:Limit} gives a comprehensive and unbiased comparison with ConvLSTM under the \textit{input-20-predict-20} setting to our method's limitations. Fig. \ref{Fig:Limit_MSE}-Fig. \ref{Fig:Limit_LPIPS} reveal the ConvLSTM outperforms the 3D-SwinSTB as the number of hidden units increases. For example, at the 15th frame, the 3D-SwinSTB experiences a 1.88\% (from 444.2705 to 452.6198) increase in MSE compared to ConvLSTM with $\{128, 64, 32\}$. Similarly, at the 10th frame, there is a 0.30\% (from 0.7787 to 0.7764) decrease in SSIM for the 3D-SwinSTB compared to ConvLSTM with $\{256, 64, 16\}$. Fig. \ref{Fig:Limit_PARAM}-Fig. \ref{Fig:Limit_AIT} reveals that ConLSTM sacrifices complexity (e.g. GFlops and ATT) to achieve higher accuracy than our 3D-SwinSTB. Note that our 3D-SwinSTB has a higher parameter quantity than ConvLSTM, but the AIT is close. Future research will focus on reducing complexity and improving accuracy.

\section{Conclusion}\label{sec7}
We have introduced a named DeepSPred spectrum prediction framework, which allows for flexible configuration of the network according to different task requirements. Based on the DeepSPred, we first have introduced a novel 3D spectrum prediction model combining a \textit{3D Patch Merging ViT-to-3D ViT Patch Expanding} symmetric flow processing strategy and a pyramid structure, denoted as 3D-SwinSTB. The 3D, shifted window and hierarchical structure of this model can accurately capture the spatiotemporal dependence, global-local feature and multi-scale feature of the spectrogram, respectively. Then, we have devised an named 3D-SwinLinear model for SOR prediction. This model directly predicts future SOR by mining features from spectrogram, achieving a commendable balance between efficiency and accuracy. To ensure the adaptability of our models across diverse spectrum services, the TL have been employed. The numerical results show that our models achieve state-of-the-art spectrum prediction performance and verify the effectiveness of the TL.
\begin{figure}
	\centerline{\includegraphics[width=68mm,height=60mm]{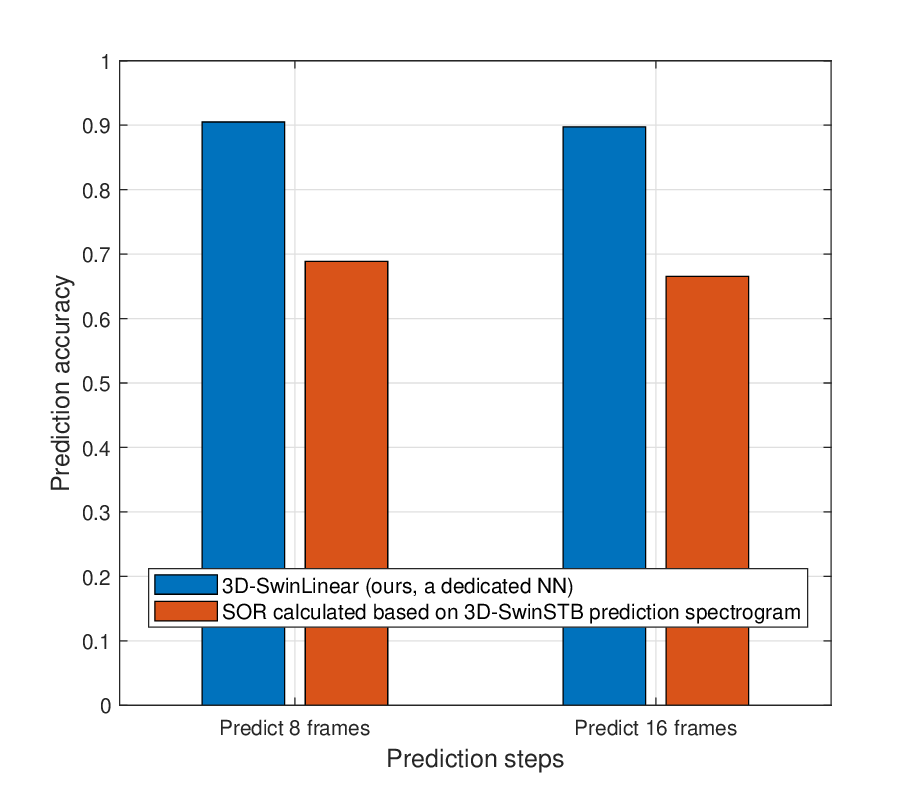}}
	\caption{Comparison of SOR prediction accuracy using the proposed 3D-SwinLinear and 3D-SwinSTB prediction spectrogram.}
	\label{Fig:dedicated}
\end{figure}

\section*{Appendix A}\label{AppA}

We grayscale the FM spectrum dataset for ablation experiments and select MSE and PSNR as evaluation metrics. As shown in Fig. \ref{Figchannel}, the MSE and PSNR of 3D-SwinSTB based on RGB spectrograms are significantly better than those of 3D-SwinSTB based on grayscale images. For example, at the tenth frame, the MSE and PSNR of the 3D-SwinSTB based on RGB spectrograms have decreased by 91.49 \% (5294.5261 $\rightarrow$ 450.7880) and increased by 210.10 \% (11.3215 $\rightarrow$ 35.0961), respectively, compared with the 3D-SwinSTB based on grayscale images. These results show: (i) RGB spectrograms provide rich spectrum usage pattern information, capturing detailed changes through the three color channels, while grayscale images lack these color details, representing only brightness; (ii) RGB spectrograms increase the diversity of the training dataset, helping the model learn various changes and features of spectrum usage patterns, thus improving generalization. In contrast, grayscale images reduce training data diversity, leading to poorer model generalization.

\section*{Appendix B}\label{AppB}

In Fig. \ref{Fig:dedicated}, we compare the prediction accuracy of the proposed 3D-SwinLinear with that of SOR calculated based on the 3D-SwinSTB-predicted spectrogram, using the FM spectrum dataset. As shown in Fig. \ref{Fig:dedicated}, the accuracy of using a dedicated network to predict SOR is superior to that of calculating SOR based on the 3D-SwinSTB-predicted spectrogram. For example, when predicting 8 frames and 16 frames, the dedicated 3D-SwinLinear achieves accuracy improvements of 31.39\% and 34.84\%, respectively, compared to the method of calculating SOR based on the 3D-SwinSTB-predicted spectrogram. This is because the former predicts SOR directly from the historical spectrogram, while the latter calculates SOR from the predicted spectrogram, which introduces additional intermediate errors due to the extra calculation steps. Furthermore, the adoption of a dedicated network does not rely on the 3D-SwinSTB, with a flexibility.

\bibliographystyle{IEEEtran}
\bibliography{myref}

\begin{thebibliography}{10}
\providecommand{\url}[1]{#1}
\csname url@samestyle\endcsname
\providecommand{\newblock}{\relax}
\providecommand{\bibinfo}[2]{#2}
\providecommand{\BIBentrySTDinterwordspacing}{\spaceskip=0pt\relax}
\providecommand{\BIBentryALTinterwordstretchfactor}{4}
\providecommand{\BIBentryALTinterwordspacing}{\spaceskip=\fontdimen2\font plus
\BIBentryALTinterwordstretchfactor\fontdimen3\font minus
  \fontdimen4\font\relax}
\providecommand{\BIBforeignlanguage}[2]{{%
\expandafter\ifx\csname l@#1\endcsname\relax
\typeout{** WARNING: IEEEtran.bst: No hyphenation pattern has been}%
\typeout{** loaded for the language `#1'. Using the pattern for}%
\typeout{** the default language instead.}%
\else
\language=\csname l@#1\endcsname
\fi
#2}}
\providecommand{\BIBdecl}{\relax}
\BIBdecl

\bibitem{sp}
G.~Pan, B.~Zhou, Q.~Wu, and D.~K. Yau, ``A 3{D} pyramid vision transformer
  learning method for spectrum prediction,'' in \emph{Proc. Wireless Commun.
  Netw. Conf (WCNC)}, submitted, Aug. 2024, pp. 1--6.

\bibitem{10106497}
D.~A. Guimarães, E.~J.~T. Pereira, and R.~Shrestha, ``Resource-efficient
  low-latency modified pietra-ricci index detector for spectrum sensing in
  cognitive radio networks,'' \emph{IEEE Trans. Veh. Technol.}, vol.~72, no.~9,
  pp. 11\,898--11\,912, 2023.

\bibitem{9424395}
M.~A. Aref and S.~K. Jayaweera, ``Spectrum-agile cognitive radios using
  multi-task transfer deep reinforcement learning,'' \emph{IEEE Trans. Wireless
  Commun.}, vol.~20, no.~10, pp. 6729--6742, 2021.

\bibitem{huang2019dynamic}
X.-L. Huang, X.-W. Tang, and F.~Hu, ``Dynamic spectrum access for multimedia
  transmission over multi-user, multi-channel cognitive radio networks,''
  \emph{IEEE Trans. Multimedia}, vol.~22, no.~1, pp. 201--214, 2019.

\bibitem{7955907}
A.~A. Khan, M.~H. Rehmani, and A.~Rachedi, ``Cognitive-radio-based internet of
  things: Applications, architectures, spectrum related functionalities, and
  future research directions,'' \emph{IEEE Wireless Commun.}, vol.~24, no.~3,
  pp. 17--25, 2017.

\bibitem{9241879}
Q.~Gao, X.~Xing, X.~Cheng, and T.~Jing, ``Spectrum prediction for supporting
  {IoT} applications over {5G},'' \emph{IEEE Wireless Commun.}, vol.~27, no.~5,
  pp. 10--15, 2020.

\bibitem{6933906}
Y.~Chen and H.-S. Oh, ``A survey of measurement-based spectrum occupancy
  modeling for cognitive radios,'' \emph{IEEE Commun. Surv. Tutor.}, vol.~18,
  no.~1, pp. 848--859, 2016.

\bibitem{8456449}
L.~Yu, J.~Chen, Y.~Zhang, H.~Zhou, and J.~Sun, ``Deep spectrum prediction in
  high frequency communication based on temporal-spectral residual network,''
  \emph{China Commun.}, vol.~15, no.~9, pp. 25--34, 2018.

\bibitem{shawel2019convolutional}
B.~S. Shawel, D.~H. Woldegebreal, and S.~Pollin, ``Convolutional {LSTM}-based
  long-term spectrum prediction for dynamic spectrum access,'' in \emph{Proc.
  27th Eur. Signal Process. Conf. (EUSIPCO)}, 2019, pp. 1--5.

\bibitem{yu2020spectrum}
L.~Yu, Y.~Guo, Q.~Wang, C.~Luo, M.~Li, W.~Liao, and P.~Li, ``Spectrum
  availability prediction for cognitive radio communications: A {DCG}
  approach,'' \emph{IEEE Trans. Cognit. Commun. Netw.}, vol.~6, no.~2, pp.
  476--485, 2020.

\bibitem{9296309}
X.~Li, Z.~Liu, G.~Chen, Y.~Xu, and T.~Song, ``Deep learning for spectrum
  prediction from spatial–temporal–spectral data,'' \emph{IEEE Commun.
  Lett.}, vol.~25, no.~4, pp. 1216--1220, 2021.

\bibitem{9664805}
X.~Ren, H.~Mosavat-Jahromi, L.~Cai, and D.~Kidston, ``Spatio-temporal spectrum
  load prediction using convolutional neural network and resnet,'' \emph{IEEE
  Trans. Cognit. Commun. Netw.}, vol.~8, no.~2, pp. 502--513, 2022.

\bibitem{liu2021swin}
Z.~Liu, Y.~Lin, Y.~Cao, H.~Hu, Y.~Wei, Z.~Zhang, S.~Lin, and B.~Guo, ``Swin
  transformer: Hierarchical vision transformer using shifted windows,'' in
  \emph{Proc. IEEE Int. Conf. Comput. Vis. (ICCV)}, 2021, pp. 10\,012--10\,022.

\bibitem{ronneberger2015u}
O.~Ronneberger, P.~Fischer, and T.~Brox, ``U-net: Convolutional networks for
  biomedical image segmentation,'' in \emph{Med. Image Comput. Comput. Ass.
  Inter. (MICCAI)}, 2015, pp. 234--241.

\bibitem{9163307}
Y.~Luo and Y.~Wang, ``A statistical time-frequency model for non-stationary
  time series analysis,'' \emph{IEEE Trans. Signal Process.}, vol.~68, pp.
  4757--4772, 2020.

\bibitem{liu2018filter}
J.~Liu, E.~Isufi, and G.~Leus, ``Filter design for autoregressive moving
  average graph filters,'' \emph{IEEE Trans. Signal Inf. Process. Netw.},
  vol.~5, no.~1, pp. 47--60, 2018.

\bibitem{8930636}
O.~Ozyegen, S.~Mohammadjafari, E.~Kavurmacioglu, J.~Maidens, and A.~B. Bener,
  ``Experimental results on the impact of memory in neural networks for
  spectrum prediction in land mobile radio bands,'' \emph{IEEE Trans. Cognit.
  Commun. Netw.}, vol.~6, no.~2, pp. 771--782, 2020.

\bibitem{7902233}
N.~Safari, C.~Y. Chung, and G.~C.~D. Price, ``Novel multi-step short-term wind
  power prediction framework based on chaotic time series analysis and singular
  spectrum analysis,'' \emph{IEEE Trans. Power Syst.}, vol.~33, no.~1, pp.
  590--601, 2018.

\bibitem{8031332}
G.~Ding, Y.~Jiao, J.~Wang, Y.~Zou, Q.~Wu, Y.-D. Yao, and L.~Hanzo, ``Spectrum
  inference in cognitive radio networks: Algorithms and applications,''
  \emph{IEEE Commun. Surv. Tutor.}, vol.~20, no.~1, pp. 150--182, 2018.

\bibitem{yu2008frequency}
C.-J. Yu, Y.-Y. He, and T.-F. Quan, ``Frequency spectrum prediction method
  based on {EMD} and {SVR},'' in \emph{Proc. 8th Int. Conf. Intell. Syst.
  Design Appl.}, vol.~3.\hskip 1em plus 0.5em minus 0.4em\relax IEEE, 2008, pp.
  39--44.

\bibitem{eltom2018cooperative}
H.~Eltom, S.~Kandeepan, Y.-C. Liang, and R.~J. Evans, ``Cooperative soft fusion
  for {HMM}-based spectrum occupancy prediction,'' \emph{IEEE Commun. Lett.},
  vol.~22, no.~10, pp. 2144--2147, 2018.

\bibitem{luo2021temporal}
S.~Luo, Y.~Zhao, Y.~Xiao, R.~Lin, and Y.~Yan, ``A temporal-spatial spectrum
  prediction using the concept of homotopy theory for {UAV} communications,''
  \emph{IEEE Trans. Veh. Technol.}, vol.~70, no.~4, pp. 3314--3324, 2021.

\bibitem{xing2013channel}
X.~Xing, T.~Jing, Y.~Huo, H.~Li, and X.~Cheng, ``Channel quality prediction
  based on {B}ayesian inference in cognitive radio networks,'' in \emph{Proc.
  IEEE INFOCOM}.\hskip 1em plus 0.5em minus 0.4em\relax IEEE, 2013, pp.
  1465--1473.

\bibitem{lin2020cross}
F.~Lin, J.~Chen, J.~Sun, G.~Ding, and L.~Yu, ``Cross-band spectrum prediction
  based on deep transfer learning,'' \emph{China Commun.}, vol.~17, no.~2, pp.
  66--80, 2020.

\bibitem{zhao2017machine}
R.~Zhao, D.~Wang, R.~Yan, K.~Mao, F.~Shen, and J.~Wang, ``Machine health
  monitoring using local feature-based gated recurrent unit networks,''
  \emph{IEEE Trans. Ind. Electron.}, vol.~65, no.~2, pp. 1539--1548, 2017.

\bibitem{9625078}
Y.~Gao, C.~Zhao, and N.~Fu, ``Joint multi-channel multi-step spectrum
  prediction algorithm,'' in \emph{Proc. IEEE Veh. Technol. Conf.}, 2021, pp.
  1--5.

\bibitem{9339826}
F.~Lin, J.~Chen, G.~Ding, Y.~Jiao, J.~Sun, and H.~Wang, ``Spectrum prediction
  based on gan and deep transfer learning: A cross-band data augmentation
  framework,'' \emph{China Commun.}, vol.~18, no.~1, pp. 18--32, 2021.

\bibitem{10064355}
G.~Pan, Q.~Wu, G.~Ding, W.~Wang, J.~Li, F.~Xu, and B.~Zhou, ``Deep stacked
  autoencoder based long-term spectrum prediction using real-world data,''
  \emph{IEEE Trans. Cognit. Commun. Netw.}, vol.~9, no.~3, pp. 534--548, 2023.

\bibitem{attention}
A.~Vaswani, N.~Shazeer, N.~Parmar, J.~Uszkoreit, L.~Jones, A.~N. Gomez,
  L.~Kaiser, and I.~Polosukhin, ``Attention is all you need,'' in \emph{Proc.
  Adv. Neural Inf. Process. Syst. (NIPS)}, 2017, pp. 5998--6008.

\bibitem{9716741}
K.~Han, Y.~Wang, H.~Chen, X.~Chen, J.~Guo, Z.~Liu, Y.~Tang, A.~Xiao, C.~Xu,
  Y.~Xu, Z.~Yang, Y.~Zhang, and D.~Tao, ``A survey on vision transformer,''
  \emph{IEEE Trans. Pattern Anal. Mach. Intell.}, vol.~45, no.~1, pp. 87--110,
  2023.

\bibitem{10039050}
G.~Pan, Q.~Wu, G.~Ding, W.~Wang, J.~Li, and B.~Zhou, ``An autoformer-csa
  approach for long-term spectrum prediction,'' \emph{IEEE Wireless Commun.
  Lett.}, vol.~12, no.~10, pp. 1647--1651, 2023.

\bibitem{9747984}
Y.~Li, T.~Yao, Y.~Pan, and T.~Mei, ``Contextual transformer networks for visual
  recognition,'' \emph{IEEE Trans. Pattern Anal. Mach. Intell.}, vol.~45,
  no.~2, pp. 1489--1500, 2023.

\bibitem{10105499}
T.~Yao, Y.~Li, Y.~Pan, Y.~Wang, X.-P. Zhang, and T.~Mei, ``Dual vision
  transformer,'' \emph{IEEE Trans. Pattern Anal. Mach. Intell.}, pp. 1--13,
  2023.

\bibitem{dosovitskiy2020image}
A.~Dosovitskiy, L.~Beyer, A.~Kolesnikov, D.~Weissenborn, X.~Zhai,
  T.~Unterthiner, M.~Dehghani, M.~Minderer, G.~Heigold, S.~Gelly \emph{et~al.},
  ``An image is worth 16x16 words: Transformers for image recognition at
  scale,'' in \emph{Proc. Int. Conf. Learn. Represent. (ICLR)}, 2020.

\bibitem{chen2021remote}
H.~Chen, Z.~Qi, and Z.~Shi, ``Remote sensing image change detection with
  transformers,'' \emph{IEEE Trans. Geosci. Remote Sens.}, vol.~60, pp. 1--14,
  2021.

\bibitem{hu2019local}
H.~Hu, Z.~Zhang, Z.~Xie, and S.~Lin, ``Local relation networks for image
  recognition,'' in \emph{Proc. IEEE Int. Conf. Comput. Vis. (ICCV)}, 2019, pp.
  3464--3473.

\bibitem{7120120}
R.~Hedjam, H.~Z. Nafchi, M.~Kalacska, and M.~Cheriet, ``Influence of
  color-to-gray conversion on the performance of document image binarization:
  Toward a novel optimization problem,'' \emph{IEEE Trans. Image Process.},
  vol.~24, no.~11, pp. 3637--3651, 2015.

\bibitem{9723467}
H.~Han, H.~Liu, C.~Yang, and J.~Qiao, ``Transfer learning algorithm with
  knowledge division level,'' \emph{IEEE Trans. Neural Netw. Learn. Syst.},
  vol.~34, no.~11, pp. 8602--8616, 2023.

\bibitem{6847217}
L.~Shao, F.~Zhu, and X.~Li, ``Transfer learning for visual categorization: A
  survey,'' \emph{IEEE Trans. Neural Netw. Learn. Syst.}, vol.~26, no.~5, pp.
  1019--1034, 2015.

\bibitem{1284395}
Z.~Wang, A.~Bovik, H.~Sheikh, and E.~Simoncelli, ``Image quality assessment:
  from error visibility to structural similarity,'' \emph{IEEE Trans. Image
  Process.}, vol.~13, no.~4, pp. 600--612, 2004.

\bibitem{zhang2018unreasonable}
R.~Zhang, P.~Isola, A.~A. Efros, E.~Shechtman, and O.~Wang, ``The unreasonable
  effectiveness of deep features as a perceptual metric,'' in \emph{Proc. IEEE
  Conf. Comput. Vis. Pattern Recognit. (CVPR)}, 2018, pp. 586--595.

\end{thebibliography}

\end{document}